# COSMIC-RAY-DRIVEN REACTION AND GREENHOUSE EFFECT OF HALOGENATED MOLECULES: CULPRITS FOR ATMOSPHERIC OZONE DEPLETION AND GLOBAL CLIMATE CHANGE

Q.-B. Lu

*Department of Physics and Astronomy and Departments of Biology and Chemistry, University of Waterloo, Waterloo, Ontario, CANADA*

**Abstract** This study is focused on the effects of cosmic rays (solar activity) and halogen-containing molecules (mainly chlorofluorocarbons—CFCs) on atmospheric ozone depletion and global climate change. Brief reviews are first given on the cosmic-ray-driven electron-induced-reaction (CRE) theory for $O_3$ depletion and the warming theory of halogenated molecules for climate change. Then natural and anthropogenic contributions to these phenomena are examined in detail and separated well through in-depth statistical analyses of comprehensive measured datasets of quantities, including cosmic rays (CRs), total solar irradiance, sunspot number, halogenated gases (CFCs, $CCl_4$ and HCFCs), $CO_2$, total $O_3$, lower stratospheric temperatures and global surface temperatures. For $O_3$ depletion, it is shown that an analytical equation derived from the CRE theory reproduces well 11-year cyclic variations of polar $O_3$ loss and stratospheric cooling, and new statistical analyses of the CRE equation with observed data of total $O_3$ and stratospheric temperature give high linear correlation coefficients ≥0.92. After the removal of the CR effect, a pronounced recovery by 20~25% of the Antarctic $O_3$ hole is found, while no recovery of $O_3$ loss in mid-latitudes has been observed. These results show both the correctness and dominance of the CRE mechanism and the success of the Montreal Protocol. For global climate change, in-depth analyses of the observed data clearly show that the solar effect and human-made halogenated gases played the dominant role in Earth's climate change prior to and after 1970, respectively. Remarkably, a statistical analysis gives a nearly zero correlation coefficient (R=−0.05) between corrected global surface temperature data by removing the solar effect and $CO_2$ concentration during 1850-1970. In striking contrast, a nearly perfect linear correlation with coefficients as high as 0.96-0.97 is found between corrected or uncorrected global surface temperature and total amount of stratospheric halogenated gases during 1970-2012. Furthermore, a new theoretical calculation on the greenhouse effect of halogenated gases shows that they (mainly CFCs) could alone result in the global surface temperature rise of ~0.6 °C in 1970-2002. These results provide solid evidence that recent global warming was indeed caused by the greenhouse effect of anthropogenic halogenated gases. Thus, a slow reversal of global temperature to the 1950 value is predicted for coming 5~7 decades. It is also expected that the global sea level will continue to rise in coming 1~2 decades until the effect of the global temperature recovery dominates over that of the polar $O_3$ hole recovery; after that, both will drop concurrently. All the observed, analytical and theoretical results presented lead to a convincing conclusion that both the CRE mechanism and the CFC-warming mechanism not only provide new fundamental understandings of the $O_3$ hole and global climate change but have superior predictive capabilities, compared with the conventional models.

## 1. Introduction

There is long and strong interest in studies of halogen-containing molecules; one of the major characteristics is their extremely strong oxidizing capability, i.e., high reactivity with an electron.[1-3] Electron-transfer reactions play important roles in many processes in physics, chemistry, atmosphere, environment, biology and medicine.[4-8] In particular, dissociative electron transfer (DET) reactions of molecules with weakly-bound, ultrashort-lived prehydrated electrons play key roles in many environmental and biological processes, ranging from the formation of the ozone hole[7,9-12] to DNA strand breaks[9,13,14]. It has been well demonstrated that DET is an extremely effective process for reactions of halogenated molecules, such as chlorofluorocarbons (CFCs) as major ozone-depleting substances, halopyrimidines and cisplatin as anti-cancer agents, with prehydrated (weakly-bound) electrons trapped on solid ice surfaces[9-11,15-27] and in liquid water[28-34], respectively. Furthermore, it was recently found that atmospheric CFCs as effective greenhouse gases may also play the major role in recent global warming observed in the late half of the 20th century.[7,35,36] Since ozone depletion and global warming are two major scientific problems of global concern, a further investigation of physical and chemical processes of halogenated molecules and their effects on the Earth climate and environment is of great significance.

Both natural and anthropogenic impacts may alter the Earth's climate and environment. First, there is long interest in studying the effects of natural factors such as solar activity and cosmic rays (CRs) on the Earth's ozone layer.[37-46] However, it has been shown that pure natural effects without involving halogenated molecules have played a limited role in forming the severe polar $O_3$ hole observed since the 1980s.[7] Thus, pure natural factors in ozone depletion will not be further discussed in this study, as one of the focuses will be on the significant and long-term (inter-annual) variation of the $O_3$ hole.

Second, with global warming observed in the late half of the 20th century, the debate on the extent to which the Sun affects the Earth's climate has become intense.[47-60] For example, a number of studies showed good correlations between solar activity and the temperature of the Earth's atmosphere on wide-range time scales of decades to centuries.[47-51] Three main mechanisms for centennial-scale solar effects on climate have been proposed.[52,55] These include changes in the energy input into the Earth's atmosphere through variations in total solar irradiance (TSI), changes in stratospheric chemistry through variations of solar UV irradiance and changes in cloud cover induced by the cosmic ray flux, which is modulated by the strength of the Sun's magnetic field. For the latte, there is considerable interest in studying the role of ions produced by CRs in the formation of aerosols and clouds in the troposphere[58-62] and in the polar stratosphere[63,64]. However, the exact role of CRs in Earth *surface* climate variations is unclear. In fact, there have been indications that the variation of solar quantities cannot explain the drastic global surface temperature rise (~0.6 °C) in the late half of the last century.[52-57] These controversies have centered on the solar influence of climate over the past three solar cycles.

Third, severe ozone depletion is known to be associated with man-made CFCs such as $CF_2Cl_2$ (CFC-12) and $CFCl_3$ (CFC-11) and the conventional understanding is the photodissociation mechanism.[65] Since its first surprising observation in 1985,[66] the large Antarctic ozone hole appears in the lower polar stratosphere in every early springtime. The Montreal Protocol has successfully phased out the production of CFCs in the world wide, and indeed the observed total halogen in the lower atmosphere has been declining since ~1994.[67-69] The photochemical models originally predicted that "*Peak global ozone losses are expected to occur during the next several years*",[67] while it turned out that one of the largest ozone holes was observed in 1998.[70] The equivalent effective stratospheric chlorine levels at mid-latitudes and Antarctica were then re-calculated to peak at ~1997 and ~2000 with respective delays of ~3 and ~6 years from the tropospheric peak, which were predicted to result in corresponding recoveries of total ozone in mid-latitudes and the Antarctic hole.[68] In contrast, however, no statistically significant recovery of ozone loss in either mid-latitudes or the polar region has been directly observed.[69] More remarkably, the largest (smallest) Antarctic $O_3$ holes were actually observed when solar activity was weakest (strongest), e.g., in 1987, 1998 and 2008 (1991, 2002 and 2012), as shown in Image 1. In fact, there has been no ozone loss observed over the Equator in the past four decades. These observations are inconsistent with the above predictions, indicating that the current photochemical theory of ozone loss is incomplete or incorrect. The ability of current atmospheric chemistry-climate models to predict the future polar ozone loss is very limited, and improving the predictive capabilities is one of the greatest challenges in polar ozone research.[71]

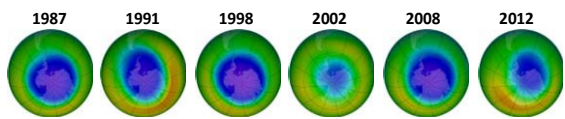

Image 1. Eleven-year cyclic October Antarctic ozone holes (satellite images credited to NASA).

Fourth, there is particularly a long history of experimental and theoretical studies of dissociative attachment (DA) of CFCs to low-energy free electrons since the 1950s.[1] It was well observed by Christophorou et al.[72,73] and Illenberger et al.[74] up to the 1970s that DA of gaseous CFC molecules to low-energy free electrons near zero eV is an extremely efficient process, and the measured cross sections of DA are far larger than those of photodissociation of CFCs. Moreover, Peyerimhoff and co-workers[75,76] first pointed out that the DA process, effectively reducing the amount of Cl released from the photolysis of CFCs, must be seen in competition to the photodissociation process and must be considered as a factor in evaluating stratospheric $O_3$ depletion. Unfortunately, however, the DA process was long thought to be insignificant for CFCs in the atmosphere due to the low free electron density detected in the stratosphere.[77,78] Then, Lu and Madey[9,10] in the late 1990s made a surprising finding that the electron-stimulated desorption (ESD) $Cl^-$ yield of CFCs is enhanced by *up to four orders of magnitude* when co-adsorption on polar $H_2O/NH_3$ ice surfaces. The electron-induced dissociation cross sections of CFCs adsorbed on polar ice surfaces were measured to be *$10^6$-$10^8$ times* the photodissociation cross sections ($10^{-20}$ cm$^2$) of gaseous CFCs. Moreover, Lu and Madey[9,10] proposed a dissociative electron transfer (DET) mechanism to explain the giant anion enhancements:

$$e_t^-(H_2O/NH_3)_n + CF_2Cl_2 \rightarrow CF_2Cl_2^{*-} \rightarrow Cl^- + CF_2Cl, \qquad (1)$$

where $e_t^-$ is a weakly-bound electron trapped in the polar ($H_2O/NH_3$) ice. In contrast to the DA process to free electrons, the DET reaction is highly effective for weakly-bound electrons that have no kinetic energies but much longer lifetimes by about two orders of magnitude than free electrons in the medium. This finding has revived the studies of electron-induced reactions of halogenated molecules including CFCs, particularly in polar media. This is evidenced by the observations first in ESD experiments[9,10,15-17], then in electron trapping experiments by Lu and Sanche[18-20], and more recently in femtosecond time-resolved laser spectroscopic measurements in polar liquids[28,29,32,33] or on ice surfaces[21,24,26]. Interestingly, a very large DET cross section up to $4 \times 10^{-12}$ cm$^2$ for $CFCl_3$ on $D_2O$ ice, measured most recently by Stähler et al.[26], is comparable to the values of $\sim 1 \times 10^{-14}$ and $\sim 6 \times 10^{-12}$ cm$^2$ for $CF_2Cl_2$ adsorbed on $H_2O$ and $NH_3$ ice respectively, originally measured by Lu and Madey[9]. The DET mechanism has also been confirmed by several theoretical studies.[23,25,27] As reviewed recently[7], it is now very well established that electron-induced dissociations of organic and inorganic chlorine, bromine and iodine containing molecules such as CFCs and HCl ($ClONO_2$) can be greatly enhanced via the DET mechanism by the presence of polar media in various phases.

Lu and Madey[9] also proposed that halogen anions (Cl$^-$) from DET reactions of halogenated molecules can be converted into active halogens to destroy ozone and this DET process should be considered as an unrecognized mechanism for the formation of the $O_3$ hole. This led to the search by Lu and Sanche[11,12] for the impact of DET reactions of halogenated molecules on the $O_3$ layer. In the stratosphere, electrons are mainly produced via atmospheric ionization by cosmic rays. The CR-driven electron-induced reaction (CRE) mechanism, as schematically shown in Fig. 1, has been proposed as an important mechanism for the formation of the $O_3$ hole.[7,9,11,12] As reviewed recently, numerous data from field measurements have provided strong evidence of the CRE mechanism.[7]

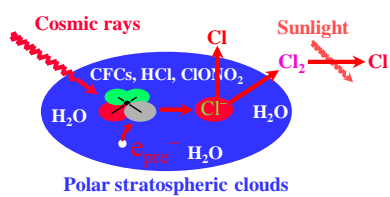

Fig. 1. The CRE mechanism[7,9,11,12]: Cosmic-ray driven electron-induced reactions of halogen-containing molecules in PSCs result in the formation of Cl$^-$ ions, which can either be rapidly converted to reactive Cl atoms to destroy the $O_3$ layer or react with other species to release photoactive $Cl_2$ and $ClNO_2$ in the winter polar stratosphere. The latter species can also then produce Cl atoms to destroy $O_3$, upon photolysis in the spring polar stratosphere.

Fifth, apart from their well-known role in $O_3$ depletion, CFCs are also long-known greenhouse (GH) gases.[79-86,68,69] In 1975, Ramanathan[79] made the first calculations that the GH effect by CFCs and chlorocarbons could lead to a rise of ~0.9 K in global surface temperature if each atmospheric concentration of these compounds is increased to 2 parts per billion (ppb). In 1987, Ramanathan et al.[80] also studied the potential climatic effects of GH gases ($CO_2$, CFCs, $CH_4$, $N_2O$, $O_3$ and others), showing that CFCs, through their indirect $O_3$-depleting effect, would have a potentially large stratospheric cooling effect, as large as that due to the $CO_2$ increase. In 1998, Ramanathan[81] further concluded that during the middle to late 20th century, the non-$CO_2$ trace gases added to the radiative heating of the planet by an amount comparable to that of $CO_2$. Without other competing factors, this total radiative heating would have warmed the planet by about 1−1.5 K by the end of last century. In 2005, Forster and Joshi[84] used various climate models to examine the role of halocarbons on stratospheric and tropospheric temperatures. They found that halocarbons (mainly CFC-12) should have contributed a significant warming of ~0.4 K at the tropical tropopause since 1950, dominating the effect of other well-mixed GH gases. They also noted that such temperature increases being "not seen" suggest that some other mechanism(s) such as stratospheric cooling due to $O_3$ loss are highly likely to be compensating for this, and as $O_3$ will likely recover in the next few decades, a slightly faster rate of warming would be expected from the net effect of halocarbons. Interestingly, Wang et al.[82,83] have noted that the spatial distribution of atmospheric opacity which absorbs and emits the long-wave radiation for non-$CO_2$ gases (CFCs) is different that for $CO_2$, for example, $CFCl_3$ is optically thin, whereas $CO_2$ is optically thick. Their simulations indicated that non-$CO_2$ GH gases provide an important radiative energy source for the Earth climate system and different infrared opacities of $CO_2$ and non-$CO_2$ GH gases can lead to different climatic effects. They concluded that it is inappropriate to use an 'effective' $CO_2$ concentration to simulate the total GH effect of $CO_2$, CFCs and other gases. However, it has generally been thought that halocarbons would play a certain but not dominant role in past and future surface temperature changes, compared with the GH effect of non-halogen gases ($CO_2$, $CH_4$ and $N_2O$).[81,84] The current atmospheric concentration of $CO_2$, ~394 parts per million (ppm), is *six orders of magnitude* higher than those of halocarbons in a few hundred parts per trillion (ppt) (far less than 1.0 ppb). Current climate models[85,86,68,69] give that the total radiative forcing $\Delta F$ of halocarbons represents only ~20% of the calculated $\Delta F$ (~1.7 W/m²) of $CO_2$, together with a small $\Delta F$ of about **-0.05 ± 0.1** W/m² due to stratospheric $O_3$ loss. It was thus concluded that $CO_2$ would play the dominant role in recent global warming.

However, it should be noted that all the above-mentioned climate models[80,81,84,85,86] assumed no *real* (*absolute*) saturation in GH effect of $CO_2$ but a *logarithmic* radiative force dependence on rising concentration. In striking contrast, recent findings strongly indicated that this assumption is invalid,[7,35,36] and the actual role of halocarbons in global climate change was re-calculated.[35] It was shown that there has been the absolute saturation, i.e., no GH effect associated with the *increasing* of concentrations of non-halogen gases, since the 1950s; atmospheric halocarbons (mainly CFCs) are most likely to cause the observed global surface temperature rise of ~0.6 °C from 1950 to 2002. Thus, global temperature is expect to reverse slowly with the projected decrease of CFCs in coming decades.[7,35] This prediction seems to be surprising, but it is actually consistent with the 2011 WMO Report.[69] The latter states that "*There have been no significant long-term trends in global-mean lower stratospheric temperatures since about 1995*", and that following an apparent increase from 1980–2000, the stratospheric water vapor amount has decreased in the past decade. These observations are "not well understood" from photochemistry-climate models,[69] but are actually consistent with the CFC warming model.[7,35] Most recently, Revadekar and Patil[36] also found the positive correlation between surface temperature and CFCs over the region of India.

It should also be noted that Müller and Grooß[87,88] recently criticized the CRE and CFC-warming theories by presenting the so-called "ACE-FTS satellite data". However, Lu[89] has pointed out that there exist serious problems with their presented data because the Canadian satellite carrying the ACE-FTS instrument has essentially *not* covered the Antarctic vortex in the presented months (especially the *winter* months when the CRE reactions are supposed to be most effective) and that their criticisms cannot stand from the scientific facts in the literature. Most recently, the pair has published a Corrigendum in one of the journals,[90] in which they state "The months for which the data were shown were not correctly indicated. ... the data do not cover this complete latitude range especially they do not extend to the South Pole". Since they now agree that their presented ACE-FTS data for the *winter* Antarctica cannot be correct, it is surprising to read their statement that "We note, however, that all conclusions of the paper remain unchanged". To discern the more data and arguments presented in the papers by Müller and Grooß[87,88], the readers should refer to the recent publication by Lu[87].

*The main purpose of this paper is to unravel the underlying mechanisms of the zone hole and recent global warming by in-depth analyses of comprehensive observed data, where the natural and anthropogenic effects will be well separated.* First, brief reviews on the CRE theory of the $O_3$ hole and the CFC theory of global warming will be given in Sections 2 and 3, respectively. Subsequently, comprehensive time-series datasets of CRs, atmospheric equivalent effective chlorine (EECl), total ozone and lower polar stratospheric temperature will be presented and analyzed in Sections 4. The natural (CR) effect will be taken out from observed data of total $O_3$ to reveal the actual effect of man-made CFCs on $O_3$ loss, and the conclusion will be examined by statistical correlation analyses of the CRE model with observed time-series data of halocarbons, CRs, $O_3$ loss and stratospheric cooling over Antarctica. This will not only examine the validity of the CRE theory but give a more true evaluation of the effectiveness of the Montreal Protocol. In Section 5, the future trend of the Antarctic $O_3$ hole will be predicted from the model calculations. Furthermore, to evaluate the potential natural effects on Earth's climate since 1850, substantial time-series datasets of global surface temperature and the most relevant solar activity indicators, including total solar irradiance (TSI), sunspot number (SSN) and CR intensity from multiple independent sources, will be examined in detail in Section 6. Time-series TSI data will include reconstructions based on models and proxies in pre-1976 and constructions based on direct measurements available since 1976. Particularly, CR measurements from as many as *ten independent stations* in different locations from the polar region to mid- and low-latitudes will be presented. To evaluate the potential human effects on Earth's climate, statistical correlation analyses of global surface temperature versus human-made $CO_2$ and halocarbons will be given in Section 7, where the solar effect will be removed from the observed temperature data. Based on observed results, a re-evaluation of the GH effects of $CO_2$ and halocarbons will be presented in Section 8. Quantitative analyses will be emphasized on the influences of humans and the Sun on global surface temperature since 1970 for several reasons. First, there is little controversy about the solar effect on the Earth's climate for the pre-1970s.[50,52] Second, the significant rise in global surface temperature occurred from 1970 to around 2000. Third, direct measurements of TSI and CFCs have only been available since the 1970s when considerable atmospheric impact ($O_3$ loss) of CFCs started to be observed. Thus, the significant anthropogenic effect on Earth's climate is expected to begin around 1970,[52] and reliable conclusions are achievable. Moreover, the future trend of global surface temperature will be presented in Section 9 and future changes of global sea level will be discussed in Section 10. Section 11 will be devoted to remarks on existing theories of ozone depletion and climate change. Finally, the conclusions will be given in Section 12. *These observations and analyses will demonstrate solid and convincing evidence of the CRE theory of the polar $O_3$ hole and the CFC-warming theory of global climate change and their superior predictive capabilities.*

## 2. The Cosmic-Ray-Driven Theory of the Ozone Hole

The main features of the CRE mechanism distinguishing from photochemical models of ozone depletion will be outlined briefly as follows.

(1) The CRE mechanism has strong *latitude* and *altitude* effects corresponding to the distribution of electrons produced by CRs in the atmosphere. Since CRs are composed of charged particles, the earth's magnetic field focuses them onto the South and North Poles, and due to atmospheric ionization, the electron production rate has a maximum at ~18 km above the ground. On the other hand, on top of the general stratosphere, the detected free electron concentration drops sharply with descending altitudes: it is ~$10^3$ electrons cm$^{-3}$ at ~85 km, and ~10 electrons cm$^{-3}$ at 60 km.[77,78] Below this height, the free electron density is too low to detect, as most free electrons are captured by atmospheric molecules (mainly $O_2$). Consequently, electron-induced decompositions of halogen-containing gases take place mainly in the upper general stratosphere at high latitudes. But the situation is drastically different in the lower polar stratosphere in winter due to the presence of PSCs.[9-12] Electrons produced by CRs can effectively be trapped in PSC ice and transferred to adsorbed halogenated molecules. As shown in Figs. 2A-D, strong *spatial* correlations between CR intensity, CFC dissociation and $O_3$ loss in the Earth's atmosphere with variations of latitude and altitude have been well observed.[7,11,89] Ozone loss occurs mainly in the polar stratosphere (Fig. 2A), and the $O_3$ hole is exactly located at the polar stratosphere at 15~18 km (Fig. 2B). Also shown in Fig. 2B is the observed $O_3$ loss over northern mid-latitudes (40°-53°N) from 1979 to 1998: apart from the expected $O_3$ loss maximum at ~40 km, there is also an $O_3$ loss peak in the lower stratosphere at the altitude of ~15 km.

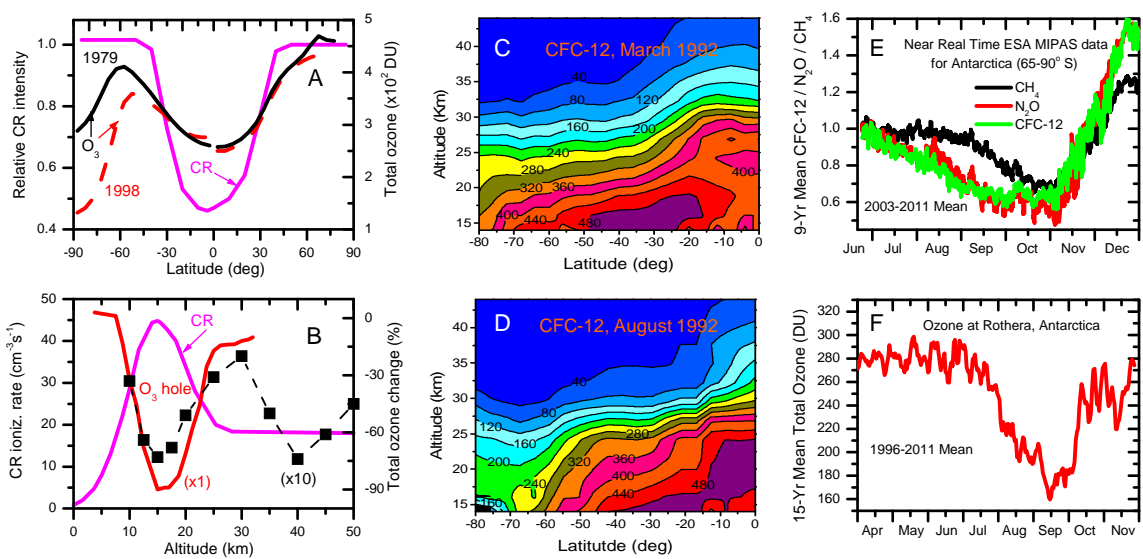

Fig. 2. Spatial and time correlations between cosmic-ray (CR) intensity, CFC dissociation and ozone depletion. A: Latitude dependences of CR intensity and monthly mean total ozone in pre-$O_3$ hole (Oct. 1979 for Antarctica and March 1979 for Arctic) and $O_3$ hole period (dashed line for Oct. 1998 for Antarctica and March 1998 for Arctic). B: Altitude dependences of the springtime $O_3$ hole over Syowa, Antarctica (the red line) and $O_3$ loss per decade from 1979 to 1998 over northern midlatitudes (40°-53°N) (the squares).[11] C and D: 27-31 March (fall) and 16-23 August (winter) 1992 $CF_2Cl_2$ levels in ppt from NASA UARS's CLEAS datasets.[7] E: 9-year mean time-series data of CFC-12, $N_2O$, $CH_4$ are averaged from the ESA's MIPAS Near Real Time (daily) satellite data in the lower Antarctic stratosphere (65-90° S) during winter months (June 23-September 30) in 2003-2011, where the data for each gas were normalized to its initial value in the beginning of winter.[89] F: The 15-year mean time series total $O_3$ data are averaged from "Real-time" daily total $O_3$ data at Rothera in Antarctica over the past 16 years (1996-2011) recorded by the British Antarctic Survey (BAS).[89] In E and F, the 2002 data were not included due to the unusual split of the polar vortex in 2002.[68]

The concentration distribution of CFCs is generally anti-correlated with the CR intensity distribution, and the decomposition of CFCs is drastically enhanced in the lower polar stratosphere during winter (Figs. 2C and D). Time-series data of $CH_4$ and CFC-12 from the NASA UARS satellite datesets during the winter season were reported earlier,[7] showing significant depletion of CFCs but not $CH_4$ in the winter months. More recently, a similar result was also observed from ESA's MIPAS Near Real Time daily satellite data of CFC-12, $N_2O$ and $CH_4$ in the lower Antarctic stratosphere (65-90° S) during the winter season (June 23-September 30) over the past ten years (2002-2011).[89] The 9-year (2003-2011) mean time-series MIPAS data are shown in Fig. 2E, which indeed shows that the CFC-12 data exhibit a variation curve that overlaps well with that of the $N_2O$ data, while the $CH_4$ data show a very different curve. It is clearly confirmed that the CFC-12 and $N_2O$ levels exhibit a similar continuous decrease since the beginning of winter, while the $CH_4$ level does not decrease until the end of August. After that, all gases show decreasing trends in September-October and then rising trends in November. Note that in September and October (the early spring), the levels of all gases ($CH_4$, $N_2O$ and CFC-12) drop in the polar lower stratosphere. This can be well explained by significant stratospheric cooling and air descending as a result of severe $O_3$ loss in the springtime lower polar stratosphere. These data have provided solid evidence of CRE (DET) reactions of CFCs and $N_2O$ but not $CH_4$ in the *winter* polar stratosphere.

The CRE mechanism can lead to the formation of reactive halogen species to destroy ozone in both the *winter* polar stratosphere in the dark and the *springtime* polar stratosphere with sunlight.[7,9] The British Antarctic Survey (BAS)'s real-time daily total $O_3$ variations at Rothera, Antarctica over the whole years since 1996 were also shown recently.[89] The 15-year mean time series total $O_3$ data are shown in Fig. 2F, which clearly shows that total $O_3$ starts to drop from a high value of about 300 DU at the beginning of July to about 220 DU (defined as the threshold of the $O_3$ hole in photochemical models[67-69]) at the middle of August and to a minimum value as low as 150 DU in September. Note that due to the lack of sunlight in the lower polar stratosphere during early and mid-winter, the significant polar $O_3$ loss in July and early August cannot be explained by photochemical models. The real-time variation of total $O_3$ (Fig. 2F) generally follows well that of CFCs or $N_2O$ (Fig. 2E) during the winter season, indicating that the CRE mechanism plays an important role in causing severe $O_3$ loss over Antarctica.

(2) The CRE model has predicted an *~11-year cyclic* variation of $O_3$ loss in the polar hole corresponding to the solar cyclic variation of the CR intensity that has an average periodicity of 11 years (varying in 9-14 years).[7,11,12] One should recall that because the oscillation amplitude of the CR intensity in 11-year CR cycles is well-known to be small, only about 10% of its mean value, the oscillation amplitude of polar stratospheric ozone will be observable only if the CRE mechanism plays a major role.[12] When the above prediction was made,[11] atmospheric chemists argued that no 11-year cyclic variations of ozone loss in the polar region would be observed.[91,92] In contrast to these arguments, high-quality ozone data obtained from NASA satellites have now confirmed a pronounced ~11-year cyclic correlation between CR intensity and 3-month average total $O_3$ data in the $O_3$ hole period over Antarctica (60-90º S).[7] Furthermore, it is also well-known that $O_3$ loss can cause a stratospheric cooling: less $O_3$ in the stratosphere implies less absorption of solar and infra-red radiation there and hence a cooler stratosphere. Thus, temperature data in the lower polar stratosphere is a direct indicator of polar $O_3$ loss. Indeed, a clear 11-year cyclic correlation between CR intensities and lower stratospheric temperatures at the Antarctic Halley station in November following the $O_3$ hole peak over the past 50 years (1956-2008) has also been found.[7]

(3) A simple quantitative expression of ozone loss due to the CRE mechanism has been found, in which total ozone loss ($\Delta[O_3]_i$) in the polar stratosphere is given by:[7]

$$\frac{\Delta[O_3]_i}{[O_3]_0} = \frac{[O_3]_i - [O_3]_0}{[O_3]_0} = -k \times [C_i] \times I_i \times I_{i-1}, \qquad (2)$$

where $[C_i]$ the equivalent effective chlorine (EECl) in the polar stratosphere, $\Delta[O_3]_i/[O_3]_0$ is the relative total $O_3$ change, $[O_3]_0$ the total $O_3$ in the polar stratosphere when $[C_i]$=0, $I_i$ and $I_{i-1}$ the CR intensities in the current and preceding years respectively, and $k$ a constant. In Eq. 2, two effects of CRs are implied: the formation of reactive halogens from CRE reactions in PSCs should be linearly proportional to $I_i$ and $[C_i]$; both CR-produced ions and $O_3$-loss-induced stratospheric cooling can affect the formation of PSCs.[7,63,64] The latter effect may be delayed by ~ 1 year.[7] It has been shown that Eq. 2 can reproduce well the observed 11-year cyclic variations of not only total $O_3$ but also stratospheric cooling over Antarctica *in the past five decades*.[7] Immediately, Eq. 2 gives $O_3$-loss maxima in 1987, 1998 and 2008 and minima in 1991, 2002 and 2012~2013, in agreement with observed data (See Image 1). More quantitative and statistical analyses of observed data in terms of Eq. 2 will be given in Section 4.

## 3. The CFC Theory of Global Warming

The explanation of blackbody radiation via the revolutionary concept of *energy quanta* put forward by Max Planck in 1900 is generally regarded as the dawn of 20th century quantum theory. Radiation from Earth's surface (after absorbing solar radiation) can approximately be treated as blackbody radiation. The radiation power in wavelength interval $d\lambda$ or frequency interval $d\nu$ is given by the Planck formula:

$$B_\lambda(T)d\lambda = \frac{8\pi hcV}{\lambda^5(e^{hc/\lambda kT}-1)}d\lambda, \quad \text{or} \quad B_\nu(T)d\nu = \frac{8\pi hV\nu^3}{c^3(e^{h\nu/kT}-1)}d\nu, \tag{3}$$

where $B_\lambda(T)$ is the power density per unit wavelength $\lambda$ and $B_\nu(T)$ is the power density per unit frequency $\nu$ (=c/$\lambda$) at the temperature $T$, where $k$ is Boltzmann's constant and $h$ is Planck's constant. Power density is defined as the radiation intensity $I$, that is, the radiation energy per unit time per unit area. Traditionally, the wavelength $\lambda_{max}$ for the $B_\lambda(T)$ maximum is given by Wien's displacement law: $\lambda_{max}$=2898 μm·K/T. Since the mean Earth's surface temperature is ~285 K, $B_\lambda(T)$ will peak at $\lambda_{max}$=10.2 μm in the infra-red (IR) wavelength range. Alternatively, $B_\nu(T)$ peaks at a wavenumber around 558.9 cm$^{-1}$, corresponding to $\lambda$=17.9 μm.

Climate researchers often argued that the IR absorption band of $CO_2$, which is at 600-770 cm$^{-1}$ ($\lambda$=13-17 μm), would center at the blackbody radiation of the Earth (see, e.g., ref. 81). However, this is actually a misconception: neither $B_\lambda(T)$ nor $B_\nu(T)$ can be compared with the atmospheric absorption (transmittance) spectrum, since they are different in physical nature and units. This paradox can be solved with the integral of either $B_\lambda(T)$ over $\lambda$ or $B_\nu(T)$ over $\nu$, which must give the same intensity $I$.[35] Here, the space radiation intensity spectrum $I(\lambda)$ of the Earth surface at T=285 K in $\lambda$=4-17 μm, together with a theoretical atmospheric transmittance spectrum, is shown in Fig. 3, which gives an intensity peak around 10 μm. There are two reasons to call the spectral region of $\lambda$=8-12 μm as the atmospheric "window". First, the unpolluted atmosphere is quite transparent in this spectral region, except for absorption by ozone at 9.6 μm; this is generally agreed by climate researchers. Second, another more critical reason is that the majority of Earth's radiation energy is emitted into space at $\lambda$=8~12 μm, where the maximum intensity of Earth's blackbody radiation is located.

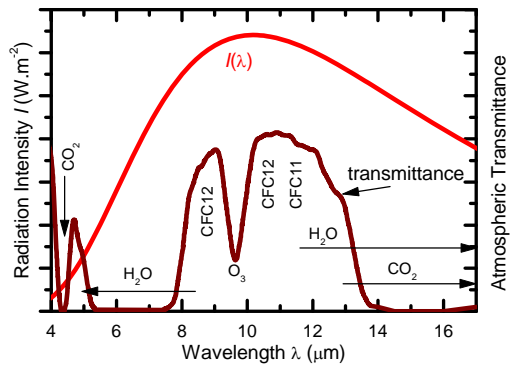

Fig. 3. Blackbody radiation intensity spectrum $I(\lambda)$ and atmospheric transmittance spectrum of the Earth.[35]

It follows that over 80% of the total radiation energy from Earth's surface and clouds is emitted into space in the $\lambda$ region of 8-12 μm. Therefore, any pollutant that strongly absorbs radiation at $\lambda$= 8-12 μm is a highly effective GH gas. Unfortunately, many halogen-containing molecules such as CFCs are not only major ozone-depleting molecules but highly effective GH gases because of their strong absorption bands at $\lambda$=8-12 μm.[35,79-84]

As also shown in Fig. 3, $CO_2$ contributes to strong absorption bands at $\lambda$=4-5 μm and 13-17 μm, while $H_2O$ (water) is the most effective absorber in the entire IR spectral range with two major bands at 5-8.3 μm and 11-17 μm. Thus, the key question is the relative importance of the GH effect of CFCs, to those of non-CFC gases (particularly $CO_2$). It is worthwhile to note that $CO_2$, $CH_4$ and $N_2O$ have high atmospheric concentrations in 392 ppm, 1.9 ppm and 327 ppb, respectively, which are $10^6$, $10^4$ and $10^3$ times those of CFCs and HCFCs in 100-500 ppt.[68,69] Although the $CH_4$ and $N_2O$ levels are lower than $CO_2$, IR absorption band strengths of $CH_4$ at 7.6 μm and $N_2O$ at 7.8 μm are 1~2 orders of magnitudes higher than that of $CO_2$ at 15 μm.[80,82] Thus, the absorption of the Earth IR radiation by these non-halogen gases is most likely to have saturated.

Indeed, the saturation in GH effect of non-halogen gases has been observed in recent studies. Contrary to the predictions of climate models[80,93,94], for example, no effects of non-halogen gases on polar ozone loss and stratospheric temperature over Antarctica since the 1950s have been found, and the global surface temperature has closely followed the total level of atmospheric CFCs.[7,35] Furthermore, changes in the Earth's GH effect can be detected from variations in the radiance spectrum of outgoing longwave radiation (OLR) at outer space, which is a measure of how the Earth radiation emits to space and carries the signature of GH gases that cause the warming effect. As first observed in a careful analysis of satellite data by Anderson et al.[95] and recently revisited by Lu[35], there exists the striking contrast between observed and $CO_2$-warming-theory predicted radiance difference between OLR spectra measured in 1970 and 1997 (spanning over the most drastic warming period). In fact, the expected strong $CO_2$ absorption band in the 600 to 800 cm$^{-1}$ region is *absent* in the observed difference spectrum. Moreover, detailed analyses by Fischer et al.[96] of high-resolution records from Antarctic ice cores showed that the $CO_2$ concentration increase by 80~100 ppm had a lag of 600~1000 years *after* the warming of the last three deglaciations, and despite strongly decreasing temperatures by about 5 °C, high $CO_2$ concentrations remained constant for thousands of years during glaciations. The results have questioned the application of the past $CO_2$-climate relation to the recent anthropogenic warming.[96] Further evidence of the real saturated GH effect of non-halogen gases and the dominant role of halocarbons in altering the Earth climate since 1970 will be shown in Sections 7 and 8.

## 4. Evaluation of the Montreal Protocol

The observations mentioned above have shown that both total $O_3$ loss and stratospheric cooling in Antarctica can be well reproduced by Eq. 2 that leads to the dependence of $O_3$ loss on the EECl level [$C_i$] and CR intensity $I_i$ only. The CR intensities have been well recorded since 1960s, showing a rising trend in the past four solar cycles.[7] This means that no sign in recovery of recorded Antarctic $O_3$ losses is most likely due to rising CR intensities, which compensates the declining EECl levels in the polar stratosphere. Thus, the *real* change of the stratospheric EECl levels can be determined by correcting measured total $O_3$ data or temperature data in the lower polar stratosphere with the CR-factor of $1/(I_i I_{i-1})$. Therefore the real effectiveness of the Montreal Protocol can be evaluated, provided that reliable data of stratospheric $O_3$ / temperatures and CRs are used.

Here, new quantitative analyses of polar stratospheric $O_3$ and cooling data in terms of the CRE mechanism are given with three improvements. First, previous studies.[7,12] used the NASA's monthly mean total $O_3$ data over the South Pole (60-90° S) directly averaged from the *original* 5° zonal-mean observed data without taking the area differences into account. This led to the absolute total $O_3$ values lower by ~30 DU, though they were mainly expressed as percentages of variations in those studies and therefore induced only a negligible artificial effect.[89] The present study uses the NASA's monthly mean total $O_3$ data over the Antarctica, re-calculated from the *area-weighted* 5° zonal-mean observed data. Second, atmospheric dynamics is known to cause large fluctuations in total $O_3$ in the polar hole from year to year.[68] To minimize this unpredictable short-term effect, a *three-point (year) adjacent averaging* is applied to observed data: $[O_3]_i = \{[O_3]_{i-1} + [O_3]_i + [O_3]_{i+1}\}/3$. This minimal processing can effectively reduce the fluctuation level of measured data. Third, *statistical correlation analyses* between the CRE equation and observed data of total $O_3$ and stratospheric cooling over Antarctica are given *for the first time*. Also note that the 2002 Antarctic stratospheric $O_3$ and temperature data, which showed unusually large deviations and could generate artificial analysis results due to the unusual split of the polar vortex,[68] are excluded in the present analyses. These improvements should lead to more accurate and reliable conclusions.

NASA's TOMS and OMI satellite datasets have so far provided the most widely used total $O_3$ data for the global and polar regions since 1979, while NOAA's ongoing surface-based observations have provided a measure of ozone-depleting chlorine- and bromine-containing gases in the lower atmosphere. Cosmic ray measurements at McMurdo (77.9° S, 166.6° E) are the only record providing *continuous* time-series CR data over Antarctica since 1960s. Here, the CR data at McMurdo and NOAA's EECl data measured in the lower atmosphere at Antarctica and mid-latitudes are plotted in Figs. 4A and B, respectively, which show that the tropospheric EECls have declined since its peak observed around 1994. Indeed, the observed tropospheric EECl over Antarctica, normalized to the 1980 value, has declined by about 24% from the peak value. The NASA's October monthly mean total $O_3$ data over the South Pole (60-90° S) and annual mean total $O_3$ in low- and mid-latitudes (65° S-65° N) in 1979-2010, as well as their 3-year average data, are shown in Figs. 4C and D, respectively. The observed data in Fig. 4C show that the Antarctic total $O_3$ decreased drastically from the end of the 1970s to 1995, following the significant rise of the halogen loading in the stratosphere. From 1995 to the present, total $O_3$ over Antarctica has exhibited pronounced 11-year cyclic oscillations. Total ozone at mid-latitudes has shown a much smaller magnitude of *continuous* decrease and clear 11-year cyclic modulations since 1979 up to the present (Fig. 4D).

The observed Antarctic and non-polar $O_3$ data after correction by the CR-factor of $1/(I_i I_{i-1})$ are shown in Fig. 4E and F, respectively, in which polynomial fits to the data give $R^2$=0.88 and 0.64 (coefficient of determination) with the probability P<0.0001 for $R^2$ = 0 (no trend). Most strikingly, Fig. 4E shows that *$O_3$ losses in the Antarctic hole have had a clear recovery since around 1995. Quantitatively, the October mean total $O_3$ loss over Antarctica has recovered by 20~25% from the peak loss at ~1995*. Comparing the $O_3$ data in Fig. 4E with the EECl data in Fig. 4A, one can clearly see that the corrected $O_3$ loss over Antarctica follows the NOAA measured EECl closely, with a short delay of only 1~2 years in the polar stratosphere. Indeed, the projected EECl with a 2-year delay has shown an approximately 23% decline from its peak at 1995-1996. *This result indicates that CFCs are indeed one of the main causes of the Antarctic $O_3$ hole, which has shown a sensitive response to the decrease in the tropospheric CFCs regulated by the Montreal Protocol.* In contrast, Fig. 4F shows no sign in recovery for $O_3$ loss at mid-latitudes. This means that a time delay of ≥10 years between the surface-measured and stratospheric EECl in mid-latitudes is required, as shown in Fig. 4B. Thus, it is obvious that the stratospheric EECl decline and associated $O_3$ recovery in mid-latitudes are significantly delayed, compared with those in the polar stratosphere.

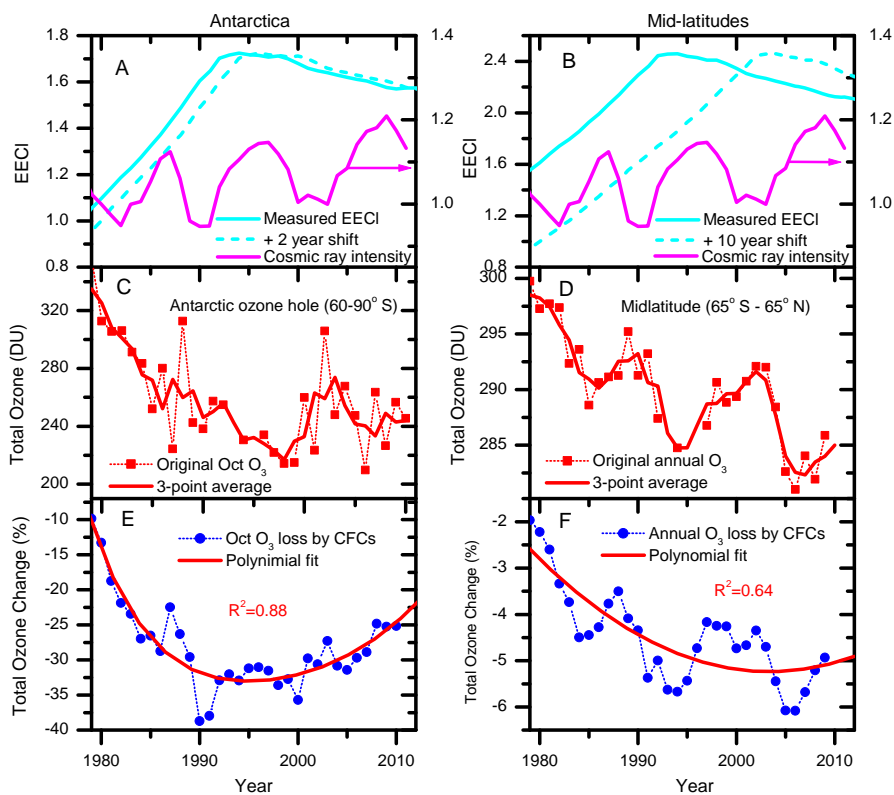

Fig. 4. Observed and corrected time-series total ozone in Antarctica (60-90° S) and mid-latitudes (65° S-65° N) during 1979-2010. A and B: NOAA's equivalent effective chlorine (EECl) data measured in the troposphere (solid lines) and projected EECl in the stratosphere (dash lines) at Antarctica and mid-latitudes, as well as CR intensities *I* measured at McMurdo, Antarctica. Both stratospheric EECl and CR data are normalized to their values in 1980. C and D: October monthly mean total $O_3$ over Antarctica and annual mean total $O_3$ at mid-latitudes, obtained from NASA TOMS N7/M3/EP/OMI satellites; also shown are the 3-point adjacent averaged smoothing. E and F: 3-point mean total $O_3$ data are corrected by the CR-factor of $1/(I_i I_{i-1})$; polynomial fits to the corrected data give coefficients of determination $R^2$ indicated and P<0.0001 for $R^2$=0 (no trend). It is clearly discovered that after the CR effect is removed, the $O_3$ hole over Antarctica has recovered by 20~25%, while no clear recovery in $O_3$ depletion at mid-latitudes has been seen.

In photochemical models, the 11-year cyclic variation of total $O_3$ in the tropics and mid-latitudes was attributed to the pure solar cycle effect: maxima in UV solar irradiance cause maxima in photochemical $O_3$ production.[68] This explanation, however, ignores another aspect of the photochemical models: maxima in UV solar irradiance would lead to maxima in activation of halogens for $O_3$ destruction. It should be noted that in the past 3 solar cycles, UV solar irradiance at the Herzberg continuum (200–242 nm) relevant to $O_3$ production varied by ~3%, which is far less than ~8% at the Schumann–Runge bands (175–200 nm) and 18% at the Schumann–Runge continuum (130–175 nm) relevant to CFC photodissociation leading to $O_3$ destruction.[97] Furthermore, photochemical models gave the maximum 11-year $O_3$ variation in the *upper* stratosphere at ~40 km, while the observed total $O_3$ cyclic variation originates mainly in the *lower* stratosphere at altitudes below 25 km for mid-latitudes and the origin remained uncertain in terms of photochemical models.[68,69] Moreover, if the solar cycle effect were significant, a similar amplitude for total $O_3$ variations over the Northern and Southern Hemispheres (NH and SH) would be expected. However, when annual total $O_3$ in 0-65°S and 0-65°N bands obtained from the same NASA TOMS/OMI satellite datasets are plotted (not shown here), one can show that the total $O_3$ over the SH shows a much more pronounced 11-year variation than that over the NH. The latter actually shows a near flat in 1983-2004. Thus, the solar cycle effect cannot be the significant cause of the observed 11-year cyclic ozone variation. A more reasonable explanation seems that the cyclic total $O_3$ variation in the extra-polar region is due to the export of CRE-driven $O_3$ loss in the polar region; CRs have an 11-year cyclic variation by about 20% from solar minima to maxima (Fig. 4A). This effect exists in both hemispheres but is larger in the SH due to the larger and more regular $O_3$ depletion in the Antarctic vortex.[68] Note also that no correction of total $O_3$ *in the polar region* by the UV solar irradiance has been proposed, as severe polar $O_3$ loss occurs mainly in the winter and early spring *lower* polar stratosphere at attitudes of 15-20 km and no solar effect is expected there.[68,69,91,92] This is also evident by the fact that polar total $O_3$ observed in the *summer* Antarctic stratosphere showed no 11-year periodic variations, indicating no solar-cycle effects on polar $O_3$ production.[7] Finally, there is no significant trend in the *long-term* variation of solar irradiance since the 1970s,[68,97] so no solar effect on the *long-term* trend of ozone in either polar or the extra-polar region is expected.

Moreover, the data in Fig. 4 are also critical to revealing the underlying mechanism for $O_3$ loss. In current context of atmospheric chemistry, the photodissociation mechanism proposed that CFCs would mainly decompose in the upper tropical stratosphere; air carrying the photoproducts (inorganic species) is then transported to the lower Antarctic stratosphere. Thus, a long delay (~6 years) from the troposphere peak was projected for the EECl to destroy $O_3$ in the Antarctic stratosphere.[68,69] The situation for the mid-latitudes of both hemispheres is different from the Antarctica primarily because it was thought that air in the mid-latitude stratosphere would have a younger mean 'stratospheric age' (~3 years) compared to air above Antarctica. As a result, halocarbons in the mid-latitude stratosphere would need less time to become degraded by UV sunlight, and hence the mid-latitude stratospheric EECl was shifted by ~3 years only from the values measured at the troposphere.[68,69] This understanding of halocarbons in the atmosphere is just opposite to the observed data shown in Fig. 4. In contrast, the CRE mechanism gives that halocarbons are mainly *in-situ* destroyed in the *polar* stratosphere and therefore the EECl to destroy $O_3$ in the polar region should be more sensitive to the change of halocarbons in the troposphere. That is, a short delay (1~2 years) is expected between the polar stratospheric EECl change and the halocarbon change observed in the troposphere. Differently due to the low electron density in the mid-latitude stratosphere (Fig. 2A), the electron-induced dissociation of halocarbons to destroy $O_3$ at mid-latitudes is much slower. In other words, halocarbons have a much longer residence time in the mid-latitude stratosphere than in the polar stratosphere. As a result, a much longer lag time from the tropospheric halocarbon change is expected for the EECl and resultant $O_3$ recovery in the mid-latitude stratosphere. This is exactly observed in the data of Fig. 4, showing strong evidence of the CRE mechanism.

To establish further the reliability of the above conclusions, statistical correlation analyses of the CRE mechanism and observed ozone data are shown in Fig. 5A-D, which plot the time-series October monthly mean and 3-month (October-December) mean total $O_3$ data observed by NASA satellites and fitted by Eq. 2 as well as total $O_3$ changes versus the product of $[C_i] \times I_i I_{i-1}$, respectively. First, as we expect, the 3-point adjacent averaging reduces the fluctuation level of total $O_3$ data significantly. Second, Figs. 5A and B show that Eq. 2 can well reproduce 11-year cyclic ozone losses in the Antarctic ozone hole. Third, Figs. 5C and D show that for the October and the 3-month average total $O_3$ data, statistical correlation coefficients -R of 0.86 and 0.92 are obtained for the linear fits to the observed data. All the statistical fits were made at a fixed 95% confidence. These results show an excellent linear correlation between total $O_3$ and the value of $[C_i] \times I_i I_{i-1}$ in the Antarctic $O_3$ hole and thus provide convincing evidence of the CRE mechanism (Eq. 2) and the reliability of the conclusions drawn from the observed ozone data corrected by $1/I_i I_{i-1}$ (Fig. 4).

As mentioned above, the temperature variation at the springtime lower polar stratosphere directly reflects real polar $O_3$ loss. Thus, it is worthwhile to show the time-series variation of the lower polar stratospheric temperature. Here, both total $O_3$ and temperatures in the lower stratosphere (100 hPa) at the BAS's Halley station (75°35' S, 26°36' W), Antarctica, in the later springs (November, immediately after the $O_3$ hole peak in October) from 1979 to 2011 are shown in Fig. 6A. First, the observed data indeed visibly show that both total ozone and lower stratospheric temperature have *11-year cyclic variations,* which can be well fitted with Eq. 2 derived from the CRE mechanism. It is particularly interesting to note that the observed temperature data have a nearly perfect fit with Eq. 2. Second, Fig. 6B shows that the lower stratospheric temperature indeed has an excellent linear dependence on total ozone, in which a high correlation coefficient R=0.93 is obtained. Third, the observed total $O_3$ and lower stratospheric temperature data after correction by the CR-factor of $1/I_i I_{i-1}$ are shown in Figs. 6C and E, in which polynomial fits to the data give $R^2$=0.93 (coefficient of determination) with the probability P<0.0001 for $R^2$ = 0 (no trend). Consistent with the NASA satellite $O_3$ data for the Antarctica (60-90° S) in Figs. 4 and 5, both total $O_3$ and lower stratospheric temperature data at Halley again show a clear recovery since ~1995. Quantitatively, the pronounced recoveries of ~25% in total $O_3$ loss and of ~22% in stratospheric cooling are clearly revealed.

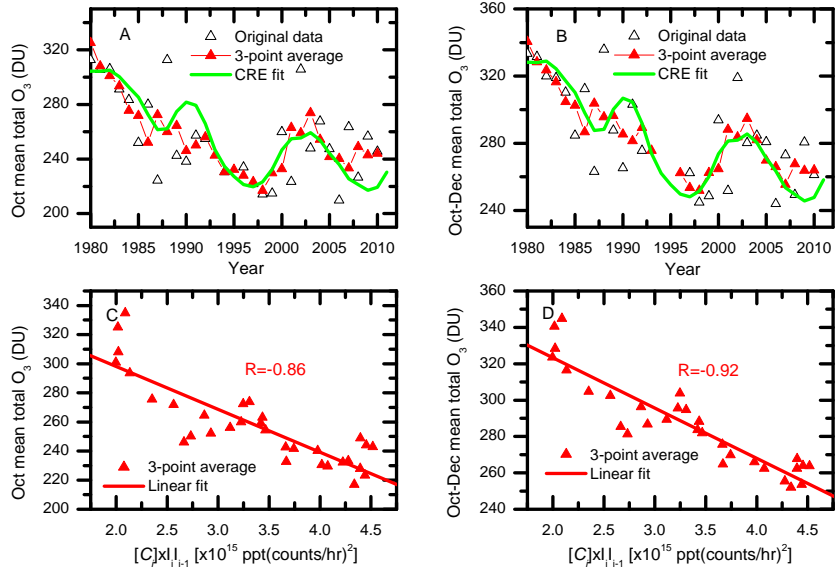

Fig. 5. Statistical analyses of the CRE mechanism and the Antarctic (60-90° S) ozone hole during 1980-2010. A and B: Time series data of October mean and 3-month (October-December) mean total $O_3$ and their 3-point adjacent averaged data, as well as the best fits by the CRE equation (Eq. 2, text). C and D: 3-point average total $O_3$ data are plotted as a function of the product of $[C_i] \times I_i I_{i-1}$, the equivalent effective stratospheric chlorine $[C_i]$ and the CR intensities $I_i$ and $I_{i-1}$ in the current and preceding years; linear fits to the data give high linear correlation coefficients -R up to 0.92 and P<0.0001(for R=0). Here, total $O_3$, EECl and CR data have the same sources as those in Fig. 4.

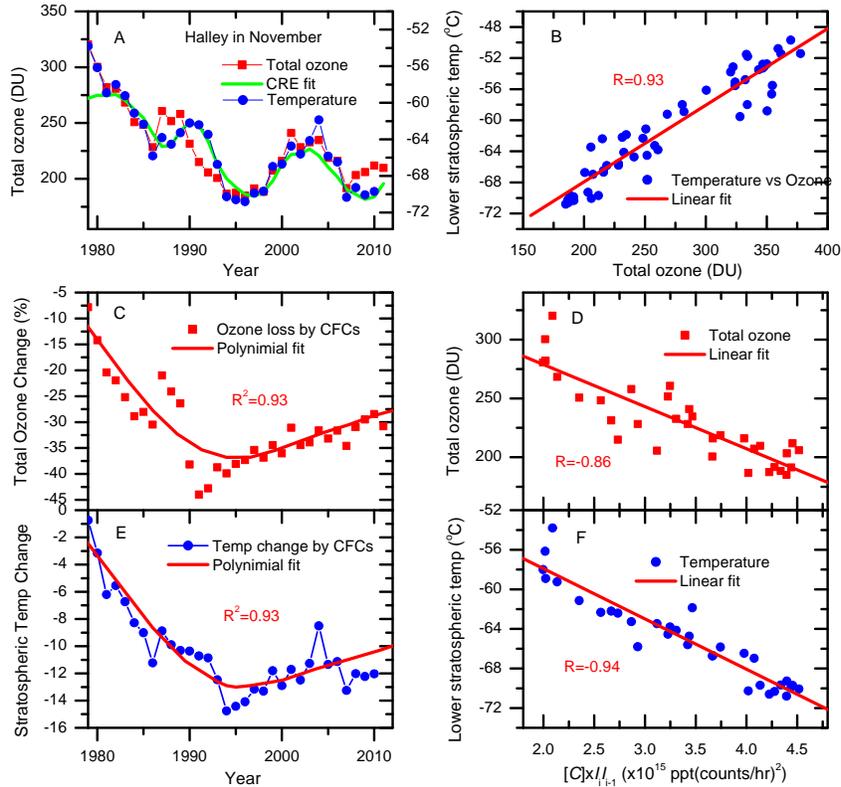

Fig. 6. Observed and corrected time-series data and statistical analyses of total ozone and lower stratospheric temperature at Halley, Antarctica in November during 1979-2011. A: Observed time-series $O_3$ and temperatures (only a minimum 3-point smoothing was applied to the observed data), as well as the best fit by Eq. 2. B: Temperatures versus total $O_3$; a linear correlation coefficient R as high as 0.93 is observed. C and E: Observed $O_3$ and temperature data are corrected by the CR-factor of $1/(I_i I_{i-1})$; polynomial fits to the correct data are also shown. It is clearly revealed that $O_3$ loss at Halley has recovered by about 25%. D and F: $O_3$ and temperature data are plotted as a function of the product of $[C_i] \times I_i I_{i-1}$, respectively; linear fits to the data are also shown. High linear correlation coefficients (-R) up to 0.94 are obtained.

Fourth, statistical correlation analyses of total $O_3$ and lower polar stratospheric temperature at Halley in terms of the CRE equation are shown in Figs. 6D and F, respectively. Again, excellent linear correlations of total $O_3$ and stratospheric temperature with $[C_i] \times I_i I_{i-1}$ variations are observed, and *high linear correlation coefficients –R up to 0.94 and P(for R=0)<0.0001 are obtained*. Similar to those shown in Figs. 5A-D, these results in Figs. 6A-F confirm that the CRE mechanism can excellently reproduce not only total $O_3$ but also $O_3$-loss-induced stratospheric cooling data in the Antarctic hole. The observed pronounced 11-year cyclic variations in total $O_3$ and lower stratospheric temperature are obviously due to the effect of CRs. These results lead to an important conclusion that both polar $O_3$ loss and lower stratospheric temperature over the past decades are well described by Eq. 2 with the equivalent effective stratospheric chlorine [*C*] and the CR intensity *I* as the only two variables. *In particular, the strikingly high linear correlation coefficient of 0.94 indicates that the long-term temperature variation in the lower polar stratosphere is nearly completely controlled by Eq. 2*. This is in striking contrast to the predictions of previous climate models that the large magnitude of stratospheric cooling due to the $CO_2$ increase would be observed,[80,93,94] even as large as that induced by $O_3$ loss.[80] *In fact, the results in Figs. 4-6 show no effect of non-halogen GH gases ($CO_2$, $CH_4$, $N_2O$, etc) on the stratospheric climate of Antarctica over the past four decades*.

In a short summary, excellent statistical correlations of total $O_3$ and polar stratospheric cooling with $[C_i] \times I_i I_{i-1}$ shown in Figs. 4-6 have strongly shown the validity of Eq. 2 to describe the Antarctic $O_3$ hole. Eq. 2 can unravel the direct effect of human-made CFCs on $O_3$ loss after the removal of the CR effect. By correcting the observed data with the CR-factor of $1/(I_i I_{i-1})$, both $O_3$ loss and stratospheric cooling in the polar $O_3$ hole have shown a clear recovering trend since around 1995, closely following the decrease of CFCs with the peak measured in the troposphere that occurred in 1993-1994. *A pronounced recovery by 20~25% of the $O_3$ hole is now clearly revealed*. This result shows the success and importance of the Montreal Protocol inhibiting the use of CFCs on a global scale.

## 5. Future Trends of the Ozone Hole

The high correlation coefficients obtained in Figs. 5 and 6 give high confidence in applying Eq. 2 to predict the future recovery of the ozone hole with the projected variations of human-made EECl and natural CRs. The Montreal Protocol has been effective in regulating ozone-depleting halogen-containing molecules, so that the EECl in the polar stratosphere is expected to continue the decreasing trend observed in the past decade.[69] The CR-intensity variation with an average periodicity of 11 years and its modulation of ~10% are well known, which can generally be expressed as[7]

$$I_i = I_{i0}\left\{1 + 10\% \sin[\frac{2\pi}{11}(i - i_0)]\right\}, \tag{4}$$

where $I_{i0}$ is the median CR intensity in an 11-year cycle. Note that $I_{i0}$ at Antarctica in the past three solar (CR) cycles has an increasing rate of ~2% per 11-year cycle. The best fit to all the observed CR data at the Antarctica (McMurdo) from 1960s-2009 yielded $I_{i0}$=8800[1+2%(i-1979)/11] ($10^2$ count/hr).[7] In ref. 7, this relationship was used to calculate future CR intensities by Eq. 4 and associated future total ozone variations over Antarctica by Eq. 2 with projected stratospheric EECl data obtained with *the assumed 6-year delay* from the tropospheric EECl data (as assumed in the 2007 WMO Report[68]). However, the observed and analyzed data shown in Figs. 4 and 6 have now shown that the stratospheric EECl data have *a lag time of only 2 years* from the tropospheric EECl data over Antarctica. Moreover, there is some uncertainty for the future change of solar activity, which regulates the future change of CRs in the coming decades. Here, we discuss three possible scenarios: A, the mean CR intensity ($I_{i0}$) in the solar cycles in the 21th century would keep the same rising rate as that in the past three solar cycles; B, $I_{i0}$ would keep nearly a constant value identical to that observed for the last 11-year cycle (2000-2011); and C, $I_{i0}$ would have a decreasing trend returning to the value in 1970. The future CR intensities calculated by Eq. 4 with these scenarios of $I_{i0}$ are shown in Fig. 7A. With these projections and projected EECl data obtained with a lag time of only 2 years from the tropospheric EECl data, the calculated October and October-December total $O_3$ over Antarctica for 1980-2080 by Eq. 2 are shown in Figs. 7B and C. The Antarctic $O_3$ hole is predicted to recover to the 1980 level around 2058(±5), depending on the variations of not only the halogen loading but also CRs in the stratosphere. This recovery is faster by about 30~40 years than that given by photochemical model calculations,[69] which predicted a recovery to the pre-1980 value by the end of this century.

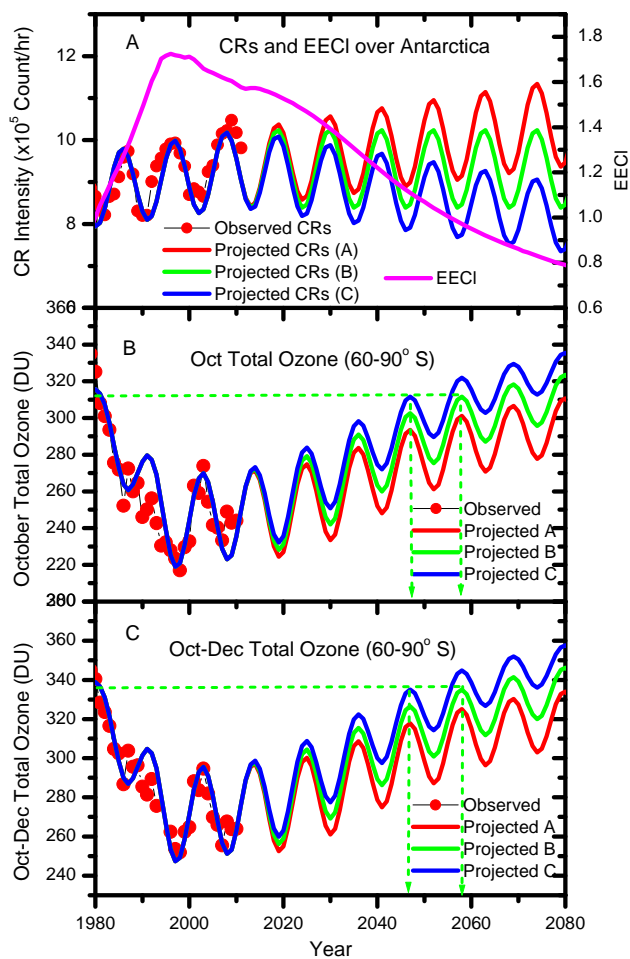

Fig. 7. Future trend of the Antarctic ozone hole. A: Observed and projected future changes of cosmic ray (CR) intensity and the equivalent effective chlorine (EECl) in the stratosphere of Antarctica, where projected EECl data were obtained with an assumed lag time of 2 years from the measured troposheric EECl data over Antarctica, observed CR data up to 2011 were obtained from the measurements at McMurdo and the future CR data were projected by Eq. 4 with the three possible scenarios of $I_{i0}$ given in the text. B and C: Calculated October monthly mean and 3-month (October-December) mean total $O_3$ over Antarctica with latitudes (60-90°S) by Eq. 2 with the observed and projected EECl and CR data shown in A; the observed total $O_3$ data (solid circles in red) from 1979 to 2010 are the same as those in Fig. 5. The projected time range for recovery of the Antarctic $O_3$ hole to the 1980 level is also indicated (arrows).

## 6. Natural Effects on Global Surface Temperature

There has been intense debate on the natural effect on Earth climate over the past 3~4 solar cycles. The controversies arise partially from the fact that direct total solar irradiance (TSI) measurements have only been available during the last three cycles and are based on a composite of many different observing satellites. As reviewed recently by Fröhlich,[57] there are some differences in constructed TSI composites in not only absolute TSI values but also time-series trends of TSI. Two representative TSI composites are the PMOD composite[54] and the ACRIM composite[51]. Both are discussed in detail in their corresponding team websites. The ACRIM composite used the data as published, whereas the PMOD composite introduced corrections of effects not considered in the original data analysis. The ACRIM composite TSI time series showed an upward trend of 0.04-0.05% per decade between consecutive solar activity minima and was proposed to account for the global temperature rise in recent decades.[51] In contrast, the PMOD TSI composite showed a downward trend of TSI at solar maxima and indicated that the measured TSI variations since 1978 are too small to have contributed appreciably to accelerated global warming over 1970-2000.[53] Krivova et al.[56] also recently re-analyzed the ACRIM composite, employing the more appropriate SATIRE-S model. Their 'mixed' ACRIM–SATIRE-S composite showed no increase in the TSI from 1986 to 1996, in contrast to the ACRIM composite; a slight decrease by approximately 0.011–0.05% was actually found though it could not be estimated very accurately. Other studies[52,54,55] have shown that the solar activity cannot have had a significant influence on global climate since ~1970, irrespective of the specific process dominant in determining Sun-climate interactions: TSI changes or solar UV changes or cloud coverage changes by CR flux variations. Lockwood and Fröhlich[55] have even concluded that all the trends in the Sun for either a direct TSI effect on climate or an indirect effect via CR-regulated cloud coverage or for a combination of the two have been in the opposite direction to that required to explain the observed rise in global temperature in the late 20$^{th}$ century.

Here, observed global surface temperatures and observed and constructed TSI composites since 1850 are shown in Fig. 8. First, it can be seen that the global temperature indeed closely followed the TSI variation up to 1970; the y-axis for TSI can be scaled so that the magnitudes of the temperature and TSI variations are similar during 1850-1970. This was actually shown previously by Hoyt and Schatten[50] and Solanki and Krivova[52], and an excellent linear correlation with coefficients of 0.83~0.97 between the TSI and the temperature was obtained.[52] This implies that the TSI values can be converted into temperatures using the linear correlation.[52] The observed data indicate that the solar effect played the dominant role in climate change prior to 1970. Second, one can see that overall the difference in the long-term TSI variations of the PMOD and ACRIM composites since 1970 has been insignificant as far as the mean TSI trend is concerned, though the TSI minima are slightly different between the two composites. In the past 4 solar cycles, the mean TSI variation has been negligible, compared with the significant rising trend in 1900-1950 and the decreasing trend in 1950-1965. It should be reasonable to assume that the dependence of global surface temperature on TSI after 1970 remains the same as that prior to 1970. The observed temperature data can then be corrected by the TSI data to remove the pure solar effect, using the linear relationship given by observed global temperature and TSI data prior to 1970. The thus corrected global temperature data are also shown in Fig. 8. It is seen that the TSI variation in either of the PMOD and ACRIM composites could not explain the significant global temperature rise from 1970 to 2002. Indeed, the solar effect has played a very minor role since 1970; the drastic global temperature rise in 1970-2002 must be due to another effect.

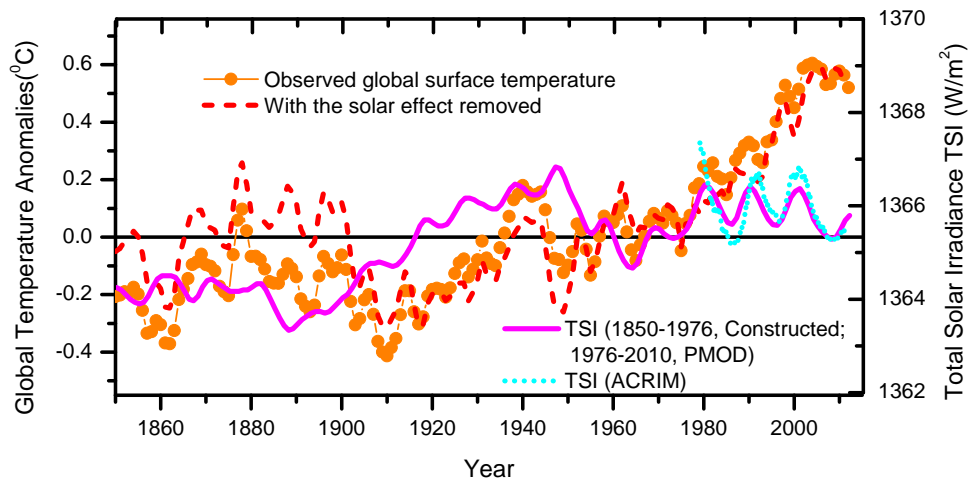

Fig. 8. Natural effects on global surface temperature. Time-series annual global (combined land and sea) surface temperature anomalies (solid points) in 1850-2012 were from the UK Met Office. Time series (annual mean) total solar irradiance (TSI) constructed based on models and proxies prior to 1976 (from Hoyt and Schatten[50]) and on direct measurements after 1976 (from ACRIM TSI by Willson[51]; PMOD TSI by Fröhlich[57]), where their absolute values are shifted to equal in the 1970s. The right y-axis range for TSI data is scaled to show their similarity to global surface temperatures prior to 1970 (see, Hoyt and Schatten[50] and Solanki and Krivova[52]). The dash curve in red is for the observed global surface temperature data with the solar effect removed (see text). Instead of using 11-point averaging,[50,52] only 3-point averaging is applied to the measured temperature and TSI data here, and the UK HadCRUT3 data are shifted by 0.15 °C to overlap with the NOAA data, to be consistent with Fig. 10.

To reach a more robust conclusion on the potential Sun's influence on climate after 1970, a more delicate analysis of other indices of solar activity may be required. Fortunately, observational records of the sunspot number (SSN) began about 300 years ago, and there has been little disagreement about the observed data of SSN. Approximately every 11 years, a maximum of solar activity is reached, with a large number of sunspots present on the solar surface. Solar cycle activity maxima are separated by minima during which only a few or no sunspots are present on the solar surface. As an indicator of solar activity, the number of sunspots is expected to have a consistent behavior with TSI. Moreover, the cosmic-ray intensity modulated by the strength of the Sun's open magnetic field is another indicator of solar activity and has been well recorded by independent detection stations in many parts of the world since the 1950s; the data are available from the Network of Cosmic Ray Stations. Thus, with combined measured data of TSI, SSN and the CR intensity from multiple sources, it is possible to obtain a reliable evaluation of the solar effect on the Earth's climate since 1970. Here, for comparison the PMOD and ACRIM TSI composites are again plotted in Fig. 9A, while the recorded time-series data of the SSN since 1900 are plotted in Fig. 9B. Fig. 9C shows the CR intensities since the 1950s from measurements *at ten stations* (McMurdo, Moscow, Apatity, Inuvik, Oulu, Kiel, Cape, Thule, Climax and Newark) at various altitudes (0-3 km) and latitudes from the polar regions to mid- latitudes. It can clearly be seen from Fig. 9B that the mean SSN had a rising trend in the first half of the 20th century and a declining trend after that, particularly in the past 3 solar cycles. The variation of the CR intensity should be anti-correlated with that of the solar activity: the solar activity minimum (maximum) corresponds to the CR maximum (minimum). As shown in Fig. 9C, indeed, the observed CR data from multiple stations show that the *overall mean CR intensity has had an increasing trend during the period of 1970 up to the present*. Clearly, the observed CR maximum at 1997 is slightly larger than that in 1986. This is opposite to the expectation from the ACRIM TSI composite[51] showing that the TSI minimum at 1997 is larger than that at 1986, while consistent with the PMOD TSI composite[53,57] and the 'mixed' ACRIM– SATIRE-S composite[56]. Overall, the combined time series data of TSI, SSN and CR intensity in Fig. 9 have unambiguously shown that the natural influence on climate is indeed in the opposite direction to that required to explain the observed rise in global temperature in the decades of 1970-2002. That is, the natural effect is negligible and the human effect must be dominant. This conclusion, consistent with recent studies,[52-57] is now clearly established by the substantial observed data shown in Fig. 9.

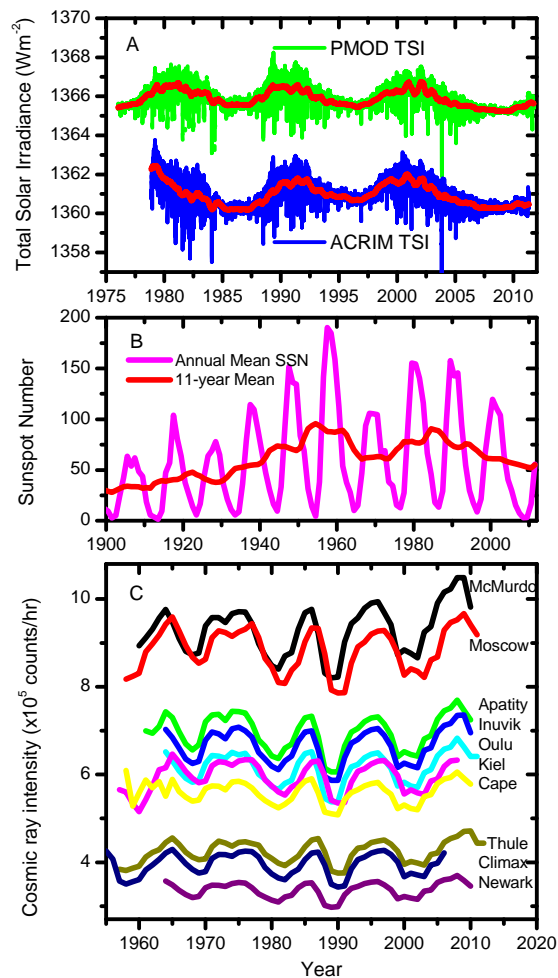

Fig. 9. Time series variations of the indicators of solar activity. A: PMOD and ACRIM TSI composites constructed from measurements since 1976/1978. B: Annual mean sunspot number (SSN) and 11-year average sunspot number (SSN) from 1900 to 2011. C: Annual Mean cosmic ray intensity data measured at ten neutron detector stations (McMurdo, Moscow, Apatity, Inuvik, Oulu, Kiel, Cape, Thule, Climax and Newark) from the 1950s to 2011.

## 7. Global Surface Temperature vs $CO_2$ and CFCs

It was recently shown that despite the continued rising of $CO_2$ since 1850, global surface temperature remained nearly constant from 1850 to ~1930; when global surface temperature was plotted versus $CO_2$ concentration, a nearly zero correlation coefficient R (=0.02) was found, that is, the global temperature was independent of the rising $CO_2$ level (285 to 307 ppm) over this period of 80 years.[35] In the present study, a more in-depth analysis of the updated data of $CO_2$, halocarbons and global surface temperature from 1850 to 2012 is presented. Figs. 10A-F divides observed data of $CO_2$, halocarbons and global surface temperature into the two periods prior to and after 1970, respectively, corresponding to the atmospheres without and with appreciable anthropogenic CFCs. Time-series data of atmospheric $CO_2$ concentration and global surface temperature from 1850 to 1970 are plotted in Fig. 10A. There was a significant temperature jump of ~0.3 °C in the short period of 1930-1940, but this can be explained well by the significant rise in solar activity,[50,52] as shown in Fig. 8. During this period of 120 years, the corrected global temperature with the solar effect removed (the same as Fig. 8) versus $CO_2$ concentration is plotted in Fig. 10B. Once again, a nearly zero correlation coefficient (R=−0.05) is obtained, indicating that the global temperatures were indeed independent of the $CO_2$ concentration rise (285-326 ppm) in 1850-1970. Time-series data of atmospheric $CO_2$ concentration, the total concentration of major atmospheric halocarbons (CFCs, $CCl_4$ and HCFCs) and global surface temperature over the four decades from 1970 to 2012 are plotted in Figs. 10C and E. In Fig. 10C, it is interesting to show that both the $CO_2$ concentrations and the total halocarbon concentrations had a nearly identical growth shape during the three decades from 1970 to~2000, while they have been drastically different since the beginning of this century. Namely, the $CO_2$ has kept the identical rising rate, whereas the total halocarbon concentration has had a turnover since ~2002. Correspondingly, it is clearly shown in Fig. 10E that global surface temperature had a linear rise from 1975 to ~2002 and has had a slowly declining trend since 2002. The 3-point smoothed temperature data in Fig. 10E also exhibits visible, 11-year cyclic small modulations, which can be attributed to the solar effect or stratospheric cooling arising from ozone loss caused by CFCs and CRs or the combination of the two effects.

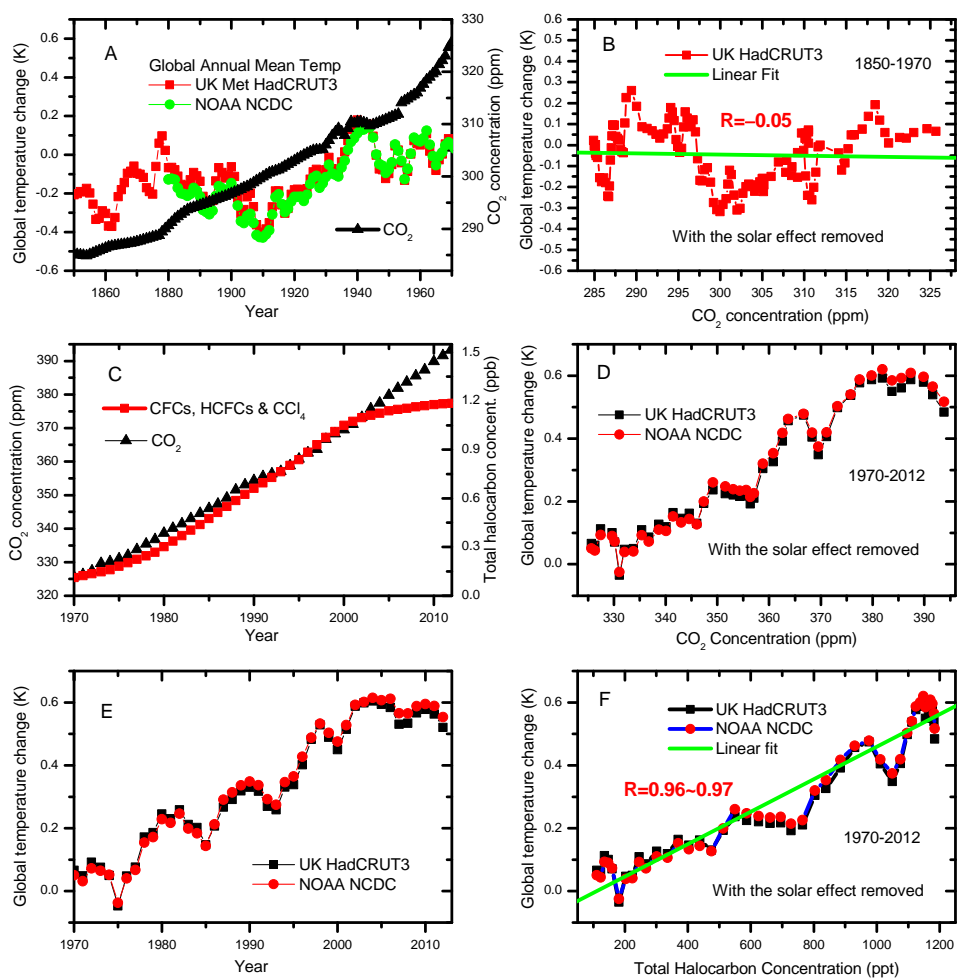

Fig. 10. Global surface temperature, $CO_2$ and halocarbons from 1850 to 2012. A and B for 1850-1970: A, time-series atmospheric $CO_2$ concentrations and global surface temperatures; B, corrected global temperature with the solar effect removed (the same as Fig. 8) versus atmospheric $CO_2$ concentration. C-F for 1970-2012: C, time-series atmospheric $CO_2$ concentrations and total concentrations of atmospheric halocarbons (CFCs, HCFCs and $CCl_4$); D, corrected global temperature with the solar effect removed versus $CO_2$ concentration; E, time-series global surface temperatures; F, corrected global temperature with the solar effect removed versus total concentration of atmospheric halocarbons. Linear fits to the data in F give nearly unit correlation coefficients as high as R=0.96-0.97 with P<0.0001 for (R=0). $CO_2$ data from 1850 to 1958 were from the Law Dome ice core analysis and data after 1958 were from direct atmospheric measurements at Mauna Loa Observatory. For halocarbon concentrations in the stratosphere, a 9-year delay from surface-based measurements was applied (see the text). Annual global surface temperatures (combined land and ocean (sea) surface temperature anomalies) were from both the UK Met Office and the US NOAA; only a minimum 3-point smoothing was applied to observed temperature data. The UK HadCRUT3 data are shifted by 0.15 °C to overlap with the NOAA data.

Most interesting are the results shown in Figs. 10D and F plotting corrected global temperature with the solar effect removed as a function of $CO_2$ concentration and total concentration of halocarbons, respectively, during 1970-2012. In Fig. 10F (and Fig. 10C), a 9-year delay in halocarbon concentrations in the stratosphere from surface-based measurements must be applied, otherwise, global surface temperature would show a sharp rise with high total halocarbon concentrations above 1100 ppt (1.1 ppb).[7,35] This delay of 9 years on the average in *global* stratospheric halocarbons was not explained previously. But it has now been well justified by the observation shown in Figs. 4 and 6 that the stratospheric EECl delays in the polar region (60°-90° S) and in non-polar latitudes (65° S-65° N) are 1~2 and ≥10 years from surface-measured EECls, respectively. Strikingly, Fig. 10F shows that corrected global surface temperature has had a nearly perfect linear dependence on the total amount of atmospheric halocarbons from the 1970 to the present. Statistically, the linear fits to the UK Met Office and NOAA data give linear correlation coefficients R as high as 0.96 and 0.97 (close to unit) and P<0.0001 for R=0, respectively. It is worthwhile to note that even without removing the solar effect from the global surface temperature data, similarly high correlation coefficients of 0.96-0.97 between global temperature and total amount of halocarbons have also been obtained.[35] This fact again indicates that the natural effect has played a negligible role in climate change since 1970, consistent with the observed results shown in Section 6. Fig. 10D shows that global temperature appeared to have a linear dependence on $CO_2$ concentration at 326-373 ppm (1970-2002), but it has been decreasing with rising $CO_2$ levels of >373 ppm (2002-present). The observed data in Fig. 10 strongly indicate that *neither the continued rises of $CO_2$ and other non-halogen GH gases nor the solar effect has played a considerable role in global temperature change, which is nearly completed controlled by the variation of atmospheric halocarbons since 1970 up to date.*

## 8. Re-evaluation of the Greenhouse Effects of $CO_2$ and Halogenated Molecules

For a GH gas, the calculation of the radiative force change $\Delta F$ with changing concentration can be simplified into an algebraic formulation that is specific to the gas. In IPCC climate models,[68,69,85,86] a *linear* dependence of the $\Delta F$ with concentration has been used for halocarbons with low concentrations (e.g., a CFC in ≤1 ppb), whereas a *logarithmic* relationship has been assumed for $CO_2$ with high concentrations (>100 ppm) so that increased concentrations have a progressively smaller warming effect. Namely, the simplified first-order expression for the radiative force $\Delta F$ of $CO_2$ is[85]:

$$\Delta F = 5.35 \times \ln(C/C_0) \qquad (5)$$

in $Wm^{-2}$, where $C$ is the $CO_2$ concentration in ppm by volume and $C_0$ is the reference concentration.

Moreover, the change in equilibrium surface temperature ($\Delta T_s$) arising from the radiative forcing $\Delta F$ of the surface-atmosphere system due to a GH gas increase can be calculated by:

$$\Delta T_s = \alpha\beta\Delta F, \qquad (6)$$

where $\alpha$ is the *climate sensitivity factor* in $K/(W/m^2)$ defined as $\alpha=dT/dF$ and $\beta$ is the *climate feedback amplification factor*. As global surface temperature rises, tropospheric water vapor increases and this represents a key positive feedback of climate change. It has been estimated that water vapor feedback acting alone approximately doubles the warming from what it would be for fixed water vapour (i.e., $\beta \approx 2$).[86] Moreover, water vapour feedback may also amplify other feedbacks such as cloud feedback and ice albedo feedback in models, which are unclear currently. In IPCC Reports[85,86], the equilibrium climate sensitivity ($\alpha\beta$) refers to the equilibrium change in global mean near-surface air temperature that would result from a doubling of the atmospheric carbon dioxide concentration ($\Delta T_{x2}$). This value was estimated to be in the range 2 to 4.5 °C with a best estimate of about 3 °C.[86] However, it also generally agreed that there exist large uncertainties in the equilibrium climate sensitivity.

*Assuming no absolute saturation in the GH effect of $CO_2$*, one could use Eq. 5 with the $CO_2$ concentration increase from 285 ppm in 1850 (the pre-industrial era) to ~390 ppm by 2010 to obtain a very large radiative force of ~1.7 W/m$^2$, as given in the 2011 WMO Report.[69] Consequently, the observed temperature increase by about 0.6 °C in the late half of last century has forced climate models to use a very low climate sensitivity, $\alpha\beta \sim 0.8$ K/(W/m$^2$).[86] The latter corresponds to a warming of ~3 K calculated by Eq. 6, with the radiative force $\Delta F$ of ~3.7 W/m$^2$ given by Eq. 5 for a doubling of $CO_2$. Although this was given in the 2007 IPCC Report,[86] unfortunately these assumptions and calculations inevitably led to the general conclusion that $CO_2$ is the major GH gas responsible for the global warming observed for 1970-2002, while halogenated molecules (mainly CFCs) are important, but by no means dominant for surface temperature changes.[80,81,84-86]

The global surface temperatures calculated by Eqs. 5 and 6 with $\alpha\beta=0.8$ K/(W/m$^2$) and measured $CO_2$ concentrations during 1850-2012 are shown in Fig. 11. Interestingly, Eq. 5 would also give an identical $\Delta F \approx 0.72$ W.m$^{-2}$ for the $CO_2$ concentration rises of 285 to 326 ppm in 1850-1970 and of 326 to 373 ppm in 1970-2002. This radiative force would cause a global surface temperature rise of about 0.6 °C given by Eq. 6, as shown in Fig. 11. It is also reasonably expected that the climate sensitivity should become lower with higher $CO_2$ concentrations.[85,86] These calculated results actually contradicts the nearly zero variation in global surface temperature with the solar effect removed during 1850-1970 and the drastic rise of ~0.6 °C during 1970-2002, as well as the observed *negative* correlation with $CO_2$ concentrations at >373 ppm for 2002 to the present. It should also be noted that if the $CO_2$-warming model were correct, the current global temperature would be at least 0.2~0.3 °C higher than the observed value (see Fig. 11). In contrast, the observed global temperature has exhibited a nearly perfect linear positive correlation ($R \approx 1.0$) with the total concentration of halogenated gases since their considerable emission into the atmosphere in the 1970s (Fig. 10F). *These observed and calculated results in Figs. 8, 10 and 11, together with those reported in refs. 7 and 35 and mentioned in Section 3, show solid evidence that the GH effect of increasing non-halogen gases has been saturated (zero) and to a good approximation, halogenated gases (mainly CFCs) have completely governed the global climate change since 1970.*

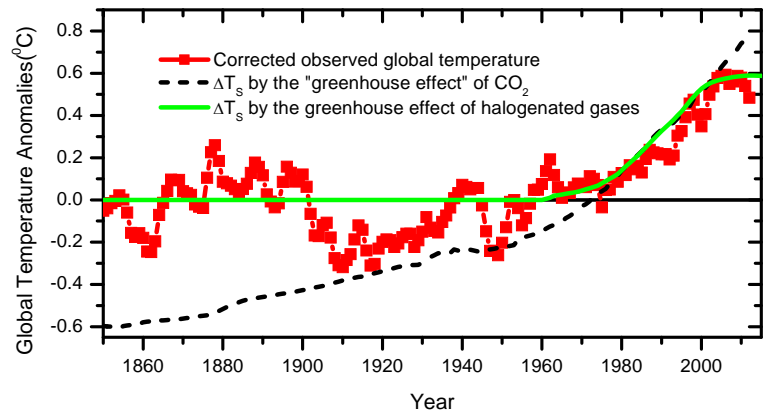

Fig. 11. Observed global surface temperatures and the equilibrium surface temperature changes ($\Delta T_s$) calculated from the $CO_2$ "greenhouse effect" given by Eqs. 5 and 6 with $\alpha\beta=0.8$ K W$^{-1}$m$^2$ [ref. 86] and from the greenhouse effect of halogen-containing molecules by Eq. 7 and 6 with $\alpha=0.9$ K W$^{-1}$m$^2$ and a feedback amplification factor $\beta=2$ [ref. 35] for 1850-2012. The observed data were obtained from the UK Met Office's combined land-surface air temperature and sea-surface temperature anomalies (a 3-point averaging was applied to the observed data) and corrected by removal of the solar effect (the same as Fig. 8).

For a GH gas at very low concentrations ($\leq 1$ ppb) such as CFCs and HCFCs, it is secure to use a simple linear dependence for the radiative forcing $\Delta F$:[68,69,85,86]

$$\Delta F = \chi C, \qquad (7)$$

where $C$ is the concentration of the GH gas and $\chi$ in $Wm^{-2}\ ppb^{-1}$ is the radiative efficiency that refers to the radiation change at the tropopause caused by a given change in $C$. *In fact, this linear dependence has now been well confirmed by the present observation of a nearly perfect linear dependence of global temperature on halocarbon concentration form 1970-2012, as shown in Fig. 10F.* The radiative efficiencies for halocarbons were calculated using radiative transfer models of the atmosphere and available in the WMO Report.[68] These $\chi$ values can be used to calculate the change $\Delta F$ as a function of changing GH concentration. In view of the *real* saturation in GH effect of non-halogen gases, simple model calculations of the global surface temperature change $\Delta T_s$ due solely to the GH effect of halocarbons were made recently with various climate sensitivities.[35] The results are also plotted in Fig. 11 to show a striking contrast to the warming model of $CO_2$. Here, a modification is made to include the 11-year cyclic small temperature modulations due to the solar/CR effect or the ozone loss effect caused by co-effects of CFCs and CRs or the combination of the two effects. Because of the observations presented in Section 6 and previous studies[52-55], showing that the natural solar effect has played only a very small part in global surface temperature change since 1950, a small modulation of 0.05 °C with an 11-year periodicity is simply assumed. This 11-year cyclic temperature variation is superposed onto the global temperature change induced by the greenhouse effect of halocarbons:

$$\Delta T_s = \Delta T_s(\text{Halocarbons}) + 0.05\cos[2\pi(i-1992)/11]. \qquad (8)$$

The resultant calculated global temperatures from 1950 to 2012 are shown in Fig. 12, together with the observed global surface temperatures since 1950. It can be clearly seen that the calculated results by Eq. 8 reproduce well the observed temperature data in 1950-2012. Here, the distinction from previous climate model calculations[80-86,68,69] is that the present calculations are based on the observed *real* saturation in the GH effect of non-halogen gases ($CO_2$, $CH_4$ and $N_2O$) and consider the GH effect of halocarbons only. *The results in Fig. 12 demonstrate that halocarbons (mainly CFCs) alone could indeed result in the observed global surface temperature rise of about 0.6 °C in 1950-2002.*

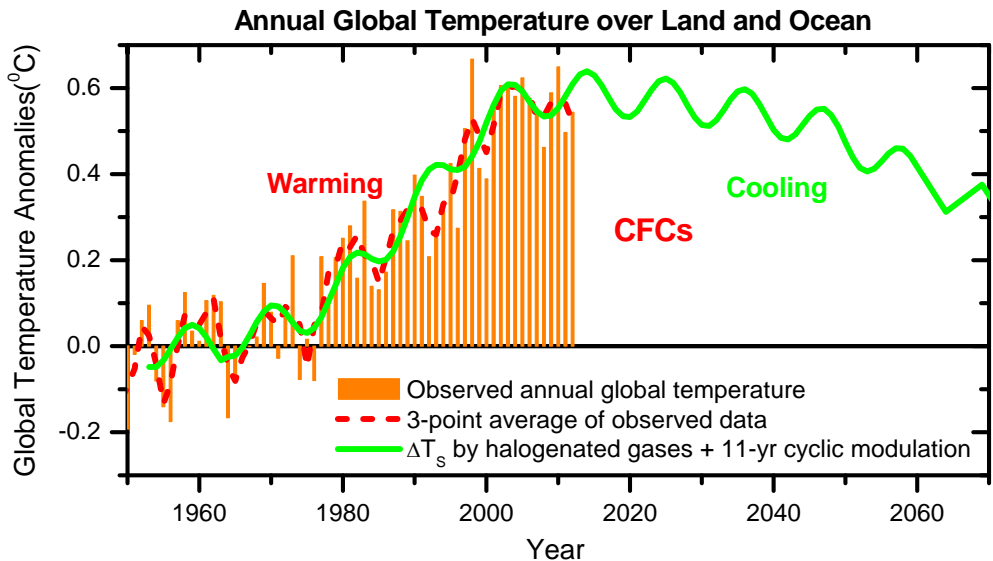

Fig. 12. Natural and human effects on global surface temperature. Observed global surface temperatures were from the UK Met Office's combined land-surface air temperature and sea-surface temperature anomalies; the red curve is 3-point average of the original observed data. The equilibrium surface temperature changes ($\Delta T_s$) (curve in green) were calculated by Eq. 8, including the contributions from the pure greenhouse effect of halocarbons (CFCs, HCFCs and $CCl_4$) obtained with a climate sensitivity factor $\alpha=0.9$ K $W^{-1}m^2$ and a feedback amplification factor $\beta=2$ [ref. 35] and from an 11-year cyclic small modulation ($\pm 0.05$ °C) due to the solar activity (CR) related effect (see text).

**9. Future Global Temperature Change**

The above Sections have strongly indicated that CFCs are the major culprit for global climate change since 1950. Thus, the future climate change will be mainly governed by the variation of halogenated molecules in the atmosphere. With the projected concentrations of major halocarbons including CFCs, HCFCs and $CCl_4$ in the future,[68,69] global surface temperatures due to halocarbons plus small 11-year cyclic modulations (within 0.05 °C) in 2013-2070 have also been calculated and shown in Fig. 12. As revealed by the observed $O_3$ data shown in Fig. 4, the decline in the total halogen burden in global stratosphere, especially in the low and middle latitudes, is much slower than that in the polar stratosphere and is significantly delayed by about one decade from that measured in the troposphere. As a result, the recovery in global surface temperature will be much slower than that in the Antarctic $O_3$ hole. The global temperature is expected to return to its mean value in 1950-1970 by the end of 21th century. Certainly, the future global temperature change might be affected by the variation of solar activity. However, there is currently no sign that the solar effect would significantly affect the Earth climate in coming decades.[98] Thus, there is good reason to predict that global surface temperature is reversing slowly with decreasing concentrations of CFCs in the atmosphere. It should also be noted that similar to the temperature trend in the past decade, the Earth in the coming decade will still be in the hottest climate over the past 150 years. However, a long-term trend of global temperature recovery (cooling) is expected for coming five to seven decades, as shown in Fig. 12.

**10. Future Global Sea Level Change**

It is well known that the change in stratospheric ozone radiatively affects surface temperature.[68,69] As shown in the Earth blackbody radiation intensity spectrum (Fig. 3), $O_3$ is an effective greenhouse gas due to its strong absorption at 9.6 μm. Currently, global stratospheric total $O_3$ in low- and mid-latitudes (65° S-65° N) has been depleted by about 6% (Fig. 4D). On the global scale, the negative radiative forcing of climate due to stratospheric $O_3$ depletion is only $-0.05 \pm 0.1$ W/m$^2$, much smaller than the positive radiative forcing due to atmospheric CFCs and HCFCs causing the depletion.[69] However, severe ozone loss in the polar hole has largely affected many aspects of the surface climate over Antarctica, even Southern hemisphere mid- and high-latitudes. These include: the southward shift of the Southern hemisphere tropospheric jet, surface winds warming over the Antarctic Peninsula and cooling over the high plateau as well as observed increases in sea-ice area averaged around Antarctica over the past decades.[69] The present observation of a faster recovery in the Antarctic ozone hole than global surface temperature, corresponding to the observation that the decline of the total halogen burden in the polar stratosphere is much faster than that in global stratosphere, means that in spite of the expected decrease in future global temperatures, the averaged melting of sea ice around the polar region will *increase* in coming decades. This will lead to an interesting phenomenon that *the global sea level will continue to rise while the global surface temperature decreases slowly in the coming one to two decades.* This trend is expected to keep until the effect of the global temperature recovery dominates over the effect of the ozone hole recovery on the sea ice area. After that, both global surface temperature and sea level are expected to drop concurrently.

**11. Remarks on Existing Theories of Ozone Depletion and Climate Change**

Current photochemistry-climate models[68,69] cannot reproduce the observed 11-year cyclic variation of polar ozone loss, nor can they capture the essential features of polar stratospheric cooling.[7] Chemistry transport models (CTMs) might partially reproduce the observed ozone, but their simulations require to use *observed* temperatures and winds and thus do not have the capability to predict future changes of the ozone hole.[89] The ability of these photochemical models including CTMs to predict future ozone hole trends is thus very limited; improving their predictive capabilities for the ozone hole is one of the greatest challenges in the ozone research community.[71] As schematically shown in Fig. 1, the CRE mechanism proposes that DET reactions of halogenated molecules can either lead to the formation of reactive Cl atoms to destroy the $O_3$ layer or react with other species to release photoactive $Cl_2$ and $ClNO_2$ in the *winter* polar stratosphere.[7,9] The latter species can also then produce Cl atoms to destroy $O_3$, upon photolysis in the spring polar stratosphere, similar to one of the steps in the photochemical models. Thus, the CRE mechanism does not exclude any possible contribution of sunlight-related photochemical processes to $O_3$ depletion. In deriving the CRE equation (Eq. 2), however, a simplification was indeed made that the photolysis of halogens in the spring polar stratosphere is not a limiting factor; this simplified CRE equation was originally aimed to give an approximate envelope of the long-term total $O_3$ variation.[7] Strikingly, it has been demonstrated in Figs. 4-6 that the CRE equation including no effects of photochemical processes reproduces well the observed data of not only polar ozone loss but stratospheric temperature change (cooling). This is in striking contrast to the photochemical models. This fact indicates that the CRE/DET reaction, rather than the photoactivation of halogen species in the gas phase, is so critical that it is indeed the limit factor leading to ozone loss in the polar ozone hole. Since its birth more than one decade ago, the CRE mechanism has well explained the observed data and shown excellent predictive capabilities.

One may be attempted to argue that since non-halogen molecules such as $CO_2$, $N_2O$ and $CH_4$ are well-known GH gases, their effects should be input into any climate models, so should the natural solar effects. This argument appears to be logical and reasonable, particularly considering the following facts. Without the atmospheric layer consisting of these GH gases even prior to the industrial revolution in the middle 19th century, the Earth surface temperature would far below the current mean temperature just suitable for living creatures. And it has no doubts that the nature (solar) factors had affected the Earth climate significantly or severely in certain periods of time, e.g, the ice ages. However, while closely looking into possible causes of recent global climate change starting around 1970, researchers must examine any conclusions based on available observations rather than on speculations or models full of assumptions and uncertainties. Executing this practice, the present study has found that the natural factors have played a negligible role in global surface temperature change since 1970, and that the concentrations of natural existing non-halogen GH gases ($CO_2$, $N_2O$ and $CH_4$) were already so high that their GH effects with rising concentrations since the modern industrial revolution have been saturated (zero), that is, their *continued increases* have made almost zero contributions to the ozone hole and global climate change. Based on these observations, the present warming theory of halogen-containing gases (mainly CFCs), including neither natural solar nor non-halogen gas effects, reproduces the observed data well and therefore shows a strong predictive capability.

## 12. Conclusions

Numerous observations have robustly shown that both the natural effect (CRs) and the human effect (CFCs) contribute to the depletion in the $O_3$ layer. According to the CRE mechanism that has been strongly supported by many data from both laboratory and field measurements, ozone loss in the polar $O_3$ hole can be well calculated with the simple CRE equation: $-\Delta[O_3]_i = k[C_i]I_iI_{i-1}$, where $[C_i]$ is the equivalent effective chlorine level in the polar stratosphere and $I_i$ the CR intensity in a particular year $i$. In this study, comprehensive time-series data sets of halocarbons, CRs, total ozone and $O_3$-loss-induced stratospheric cooling have given excellent quantitative and statistical analysis results consistent with the CRE mechanism. After the removal of the CR effect, a pronounced recovery by 20~25% of the Antarctic $O_3$ hole since ~1995 is discovered, while no sign in recovery of $O_3$ loss in mid-latitudes has been observed. The polar $O_3$ hole has shown a sensitive response to the decline in total halogen burden in the low troposphere since 1994 due to the regulation by the Montreal Protocol. The CRE equation has well reproduced 11-year cyclic variations of the Antarctic $O_3$ hole and the associated stratospheric cooling and significantly improved our predictive capabilities for future polar ozone loss.

Furthermore, the substantial combined data of total solar irradiance, the sunspot number and cosmic rays from multiple measurements have unambiguously demonstrated that the natural factors have played a negligible effect on Earth's climate since 1970. Moreover, in-depth analyses of time-series data of $CO_2$, halogen-containing molecules and global surface temperature have shown solid evidence that the GH effect of increasing concentrations of non-halogen gases has been saturated (zero) in the observed data recorded since 1850. In particular, a statistical analysis gives a nearly zero correlation coefficient (R=-0.05) between $CO_2$ concentration and the observed global surface temperature corrected by the removal of the solar effect during 1850-1970. In contrast, a nearly perfect linear correlation with coefficients of 0.96-0.97 is obtained between corrected or uncorrected global surface temperature and total level of stratospheric halogenated molecules from the start of considerable atmospheric CFCs in 1970 up to the present. These results strongly show that the recent global warming observed in the late 20[th] century was mainly due to the GH effect of human-made halogen-containing molecules (mainly CFCs). Moreover, a refined calculation of the GH effect of halogenated molecules has convincingly demonstrated that they (mainly CFCs) alone accounted for the global temperature rise of about 0.6 °C in 1970-2002. Owing to the effectiveness of the Montreal Protocol, the globally mean level of halogen-containing molecules in the stratosphere has entered a very slow decreasing trend since 2002. Correspondingly, a very slow declining trend in the global surface temperature has been observed. It is predicted that the success of the Montreal Protocol will lead to a long-term slow return of the global surface temperature to its value in 1950-1970 for coming 50-70 years if there is no significant emission of new GH species into the atmosphere.

In summary, the observed data have convincingly shown that CFCs are the major culprit not only for $O_3$ depletion via conspiring with cosmic rays but also for global warming by ~0.6 °C during 1970~2002. The successful execution of the Montreal Protocol has shown its fast effectiveness in controlling the $O_3$ hole in the polar region and a slow cooling down of the global surface temperature. The $O_3$ loss in the polar region is estimated to recover to its 1980 value by 2058, faster than recently expected from photochemical model simulations,[68,69] while the return (lowering) of global surface temperature will be much slower due to the slow decline of the stratospheric halogenated molecules in low and mid latitudes. This leads to an interesting prediction that global sea level will continue to rise in coming 1~2 decades until the global temperature recovery dominates over the $O_3$ hole recovery. After that, both global surface temperature and sea level will drop concurrently. It should also be noted that the mean global surface temperature in the next decade will keep nearly the same value as in the past decade, i.e., "the hottest decade" over the past 150 years. This, however, does not agree with the warming theory of $CO_2$. If the latter were correct, the current global temperature would be at least 0.2~0.3 °C higher than the observed value. Actually a slow cooling trend has begun.

This study also shows that correct understandings of the basic physics of cosmic ray radiation and the Earth blackbody radiation as well as their interactions with human-made molecules are required for revealing the fundamental mechanisms underlying the ozone hole and global climate change. When these understandings are presented with observations objectively, it is feasible to reach consensuses on these scientific issues of global concern. Finally, this study points out that humans are mainly responsible for the ozone hole and global climate change, but international efforts such as the Montreal Protocol and the Kyoto Protocol must be placed on firmer scientific grounds. This information is of particular importance not only to the research community, but to the general public and the policy makers.

**Acknowledgements.** The author is greatly indebted to the following Science Teams for making the data used in this study available: NASA TOMS and OMI Teams (especially Drs. P. K. Bhartia, R. McPeters, A. Krueger, J. Herman), NASA UARS's CLAES Team (especially Drs. A. E. Roche and J. B. Kumer), the British Antarctic Survey's Ozone Team (Dr. J. D. Shanklin), the University of Oxford's MIPAS team (Dr. Anu Dudhia), the Bartol Research Institute's Neutron Monitor Team (Dr. John W. Bieber) and the Network of Cosmic Ray Stations, the UK Met Office Hadley Centre, the US NOAA National Climatic Data Center (NCDC) and Global Monitoring Division, the US DOE Oak Ridge National Laboratory (ORNL)'s Carbon Dioxide Information Analysis Center (CDIAC), the Royal Observatory of Belgium's Solar Influences Data Analysis Center (SIDC), the Swiss PMOD / WRC team (Dr. Claus Fröhlich) and the US NASA's ACRIM team (Dr. Richard C. Willson). This work is supported by the Canadian Institutes of Health Research and Natural Science and Engineering Research Council of Canada.

**Data Sets (Sources).** NASA TOMS (N7/M3/EP) and OMI ozone satellite datasets were obtained from http://toms.gsfc.nasa.gov. The $CH_4$ and $CF_2Cl_2$ data in the stratosphere were obtained from the NASA Goddard Space Flight Center (GDFC) CLAES datasets. The $O_3$ and lower stratospheric temperature data at Rothera and Halley, Antarctica were from the British Antarctic Survey (BAS) (http://www.antarctica.ac.uk/met/jds/ozone/), credited to Dr. Jon D. Shanklin. Near Real Time satellite data of CFC-12, $N_2O$ and $CH_4$ were from The University of Oxford's MIPAS team (Dr. Anu Dudhia) (http://www.atm.ox.ac.uk/group/mipas/L2OXF/trend/). Cosmic ray data at McMurdo were from the Neutron monitors of the Bartol Research Institute http://neutronm.bartol.udel.edu). Comic ray data for other detecting stations were from the Network of Cosmic Ray Stations (http://cr0.izmiran.rssi.ru/common/links.htm). The sunspot number data were from the Royal Observatory of Belgium's Solar Influences Data Analysis Center (http://sidc.oma.be/). The PMOD TSI data were from the Swiss Science Foundation's PMOD / WRC team (Dr. Claus Fröhlich) (http://www.pmodwrc.ch/pmod.php?topic=tsi/composite/SolarConstant), while the ACRIM TSI data were from the US NASA's ACRIM team (Dr. Richard C. Willson) (http://www.acrim.com/). Global surface temperature data were from two sources, namely the UK Met Office Hadley Centre: the HadCRUT3 dataset, combined land-surface air temperature and sea-surface temperature anomalies (http://hadobs.metoffice.com/), and the US NOAA National Climatic Data Center (NCDC): the dataset of annual global (combined land and ocean temperature) anomalies (http://www.ncdc.noaa.gov/oa/climate/research/anomalies/index.html). $CO_2$ data from 1850 to 1958 were from the US DOE Oak Ridge National Laboratory (ORNL)'s Carbon Dioxide Information Analysis Center (CDIAC)'s Law Dome DE08, DE08-2, and DSS ice core measurements (http://cdiac.ornl.gov/trends/co2/lawdome-data.html); for the years without recorded data, a linear extrapolation was used to obtain the $CO_2$ data between two closest recorded data points. $CO_2$ data after 1958 were from direct atmospheric measurements at Mauna Loa Observatory, Hawaii by the NOAA's Global Monitoring Division (GMD) (http://www.esrl.noaa.gov/gmd/ccgg/trends/). Halocarbon concentrations and EECl values were from the WMO Reports[68,69] or from the NOAA's GMD.